# Smart Learning to Find Dumb Contracts
## (Extended Version)

Tamer Abdelaziz[†]
tamer@comp.nus.edu.sg
(†) National University of Singapore
Singapore

Aquinas Hobor[‡,†]
a.hobor@ucl.ac.uk
(‡) University College London
London, United Kingdom

## Abstract

We introduce *Deep Learning Vulnerability Analyzer (DLVA)*, a vulnerability detection tool for Ethereum smart contracts based on powerful deep learning techniques for sequential data adapted for bytecode. We train DLVA to judge bytecode *even though the supervising oracle, Slither, can only judge source code*. DLVA's training algorithm is general: we "extend" a source code analysis to bytecode without any manual feature engineering, predefined patterns, or expert rules. DLVA's training algorithm is also robust: it overcame a 1.25% error rate mislabeled contracts, and—the student surpassing the teacher—found vulnerable contracts that Slither mislabeled. In addition to extending a source code analyzer to bytecode, DLVA is much faster than conventional tools for smart contract vulnerability detection based on formal methods: DLVA checks contracts for 29 vulnerabilities in 0.2 seconds, a 10–1,000x speedup compared to traditional tools.

DLVA has three key components. First, *Smart Contract to Vector* (SC2V) uses neural networks to map arbitrary smart contract bytecode to an high-dimensional floating-point vector. We benchmark SC2V against 4 state-of-the-art graph neural networks and show that it improves model differentiation by an average of 2.2%. Second, *Sibling Detector* (SD) classifies contracts when a target contract's vector is Euclidian-close to a labeled contract's vector in a training set; although only able to judge 55.7% of the contracts in our test set, it has an average Slither-predictive accuracy of 97.4% with a false positive rate of only 0.1%. Third, *Core Classifier* (CC) uses neural networks to infer vulnerable contracts regardless of vector distance. We benchmark DLVA's CC with 10 "off-the-shelf" machine learning techniques and show that the CC improves average accuracy by 11.3%. Overall, DLVA predicts Slither's labels with an overall accuracy of 92.7% and associated false positive rate of 7.2%.

Lastly, we benchmark DLVA against nine well-known smart contract analysis tools. Despite using much less analysis time, DLVA completed every query, leading the pack with an average accuracy of 99.7%, pleasingly balancing high true positive rates with low false positive rates.

## 1 Introduction

We developed the Deep Learning Vulnerability Analyzer (DLVA) to help developers and users of Ethereum smart contracts detect security vulnerabilities. DLVA uses deep learning (neural networks) to analyze smart contracts. DLVA has no built-in expert rules or heuristics, learning which contracts are vulnerable during an initial training phase.

We focus on Ethereum since it has the largest developer and user bases: about 6,000 active monthly developers and 800,000 active wallets [15]. Ethereum distributed applications (dApps) target domains including financial services, entertainment, and decentralized organizations. Unfortunately, being computer programs, smart contracts are prone to bugs.

Bugs occur in smart contracts for many reasons, *e.g.* the semantics for Ethereum Virtual Machine (EVM) instructions is more subtle than is typically understood [64]. Poor software engineering techniques, *e.g.* widespread copying/pasting/modifying [29, 42] lead to rapid propagation of buggy code.

Since smart contract bytecode—and for about a third of the contracts, source code—is public, attackers can analyze a smart contract's code for vulnerabilities [48, 84]. With some contracts controlling digital assets valued in the hundreds of millions of US dollars, the motivation to attack is significant. Smart contract bugs have caused major financial losses, with various bugs costing tens or even hundreds of millions of US dollars [66, 70]. Unlike with conventional financial systems, users typically have no recourse to recover losses.

Approximately two-thirds of Etherum contracts do not have source code available, but most previous vulnerability analyzers require (or at least meaningfully benefit from) source code availability. *DLVA works directly on bytecode*. Moreover, most previous tools require significant time to analyze contracts, especially as the contracts get longer. *DLVA checks a typical contract in 0.2 seconds*, 10-1,000+ times faster than competitors, enabling vulnerability detection at scale.

We trained DLVA using contracts labeled by the Slither [25] static analyzer. Slither is state of the art but requires source code, and so can only label 32.6% of the contracts in our data set. *Although Slither can only label source code, we train DLVA to judge bytecode, thus "extending" a source code analyzer to bytecode.* Slither taught DLVA 29 vulnerabilities for long contracts (750+ opcodes) and 21 for shorter contracts.

Figure 1 benchmarks DLVA against nine competitors. DLVA is on the far right. We use bar-and-whiskers where star $\star$ represents the mean and plus $+$ represents outliers. Our average Completion Rate (*i.e.*, the percentage of contracts for which a tool produces an answer, the higher the better) is 100.0%. Our average accuracy is 99.7% (the higher the better), with a True Positive Rate (*i.e.*, detection rate; the higher the better) of 98.7% and a False Positive Rate (*i.e.*, false alarm rate; the lower the better) of 0.6%. Our average analysis time per contract (the graph is in log scale, lower better) is 0.2 seconds. We discuss Figure 1 in more detail in §4.4.

Smart learning pays off: DLVA beats Slither on every statistic except for TPR (where it lags by 0.7%). Recall also that Slither requires source whereas DLVA needs only bytecode.

Our main contributions are as follows:

1. §3.2, 3.3.1, 4.2 We develop a Smart Contract to Vector (SC2V) engine that maps smart contract bytecode into a high-dimensional floating-point vector space. SC2V uses a mix of neural nets trained in both unsupervised and supervised manners. We use Slither for supervision, labeling each contract as vulnerable or non-vulnerable for each of the 29 vulnerabilities we handle. *We provide no expert rules or other "hints" during training.* We evaluate the SC2V engine against four state-of-the-art graph neural networks and show it is 2.2% better than the average competitor and 1.2% better than the best.

2. §3.5, 4.3.1 Our Sibling Detector (SD) classifies contracts according to the labels of other contracts Euclidian-nearby in the vector space. Our SD is highly accurate, showing the quality of SC2V: on the 55.7% of contracts in our test set that it can judge, it has an accuracy (to Slither) of 97.4% and an associated FPR of only 0.1%.

3. §3.3.2, 4.3.1, 4.2 We design the Core Classifier (CC) of DLVA using additional neural networks, trained in a supervised manner using the same labeled dataset as SC2V. On the "harder" 44.3% of our test set, the CC has an accuracy (to Slither) of 80.0% with an associated FPR of 21.4%. We evaluate the CC against ten off-the-shelf machine learning methods and show that it beats the average competitor by 11.3% and the best by 8.4%.

4. §3, 4.3.1 DLVA is the combination of SC2V, SD, and CC. This whole is greater than its parts: DLVA judges every contract in the test set, with an average accuracy (to Slither) of 87.7% and FPR of 12.0%.

5. §3.6, 4.3.2 Small contracts are simpler than larger ones. We tweak our design to better handle such contracts and retrain. On small contracts, DLVA has an average accuracy (to Slither) of 97.6% with a FPR of 2.3%.

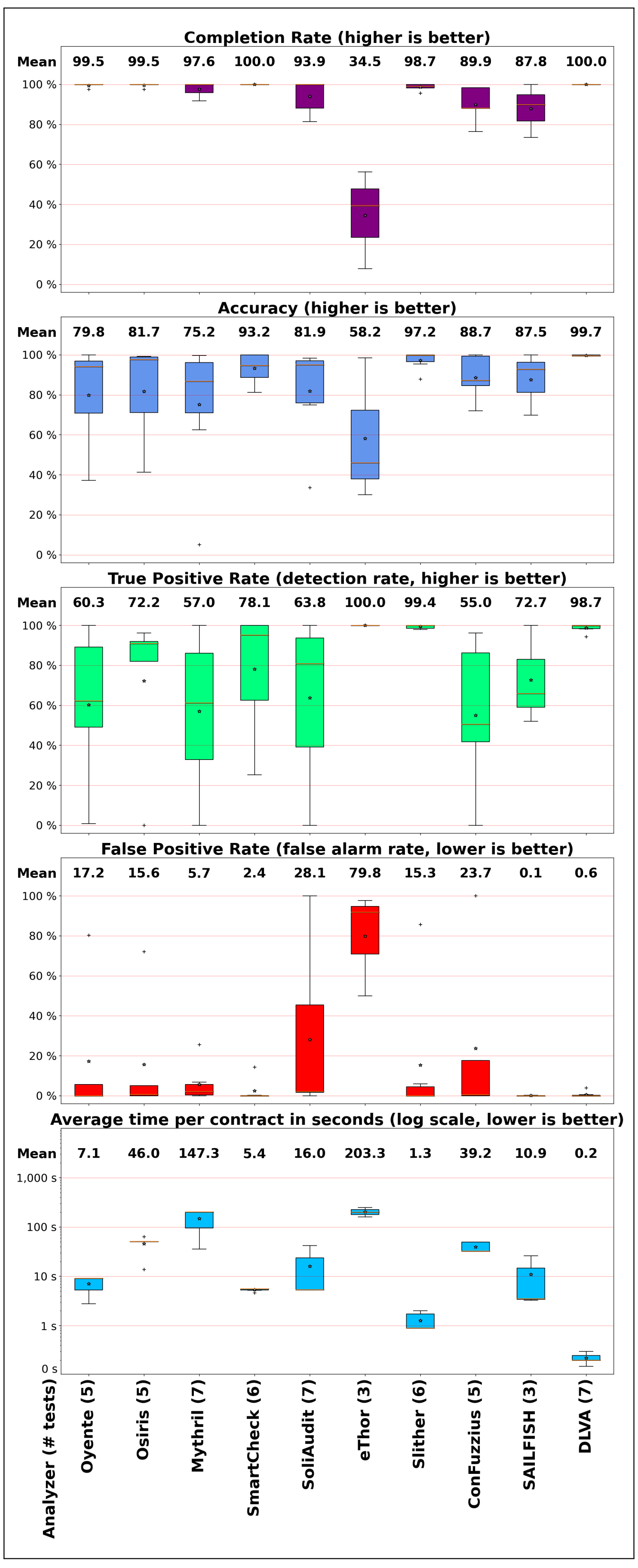

Figure 1: DLVA vs. alternatives on the $Elysium_{benchmark}$ [2], $Reentrancy_{benchmark}$ [6], and $SolidiFI_{benchmark}$ [8]; star $\star$ indicates the mean; plus $+$ indicates outliers

6. §3, 4.3.3 Accordingly, DLVA's overall accuracy (average of large and small) is 92.7% with a FPR of 7.2%.

7. §4.1, 4.4 We propose and evaluate six datasets to benchmark DLVA and its components. As presented in Figure 1, we benchmark DLVA against eight static analyzers and the machine learning based SoliAudit; SoliAudit and the static analyzer ConFuzzius also use fuzzing.

In addition to our main contributions, the rest of the paper is straightforwardly organized with background material in §2, a discussion of related work in §5, and conclusion in §6.

**Supplemental material** This paper is the extended version of a conference paper [7]. The first 18 pages are identical to the conference version, except for page numbers; this paragraph; reference [7]; and Appendix A, which overviews the extended appendices (C–L). These begin on page 19 of this paper. When we reference a Figure/Table in the extended appendix, it is prefaced by "Appx," *e.g.* Table Appx.12.

**DLVA availability and ethical considerations** Any vulnerability analyzer can be used with ill intent. Blockchains are tricky for responsible disclosure [10]. Not only are attackers incentivized to find and attack vulnerabilities, but due to the pseudonymous nature of the blockchain, it is impossible to quietly inform participants of discovered vulnerabilities.

On the other hand, since DLVA requires labelled data sets to train, none of our detected vulnerabilities are "zero-day." Moreover, honest actors benefit from DLVA too: everyone wants to know if the contracts they use are vulnerable.

On balance, the community benefits from access to DLVA, and so like other smart contract vulnerability analyzers [48, 72, 54, 71, 47, 62, 25, 73, 12], we will release DLVA.

DLVA is available for download from **`https://bit.ly/DLVA-Tool`** (see "README.md"). The instructions contain explanation for how to analyze a single contract or a batch of contracts. It takes 1–2 minutes to load the models into memory (≈2 seconds per model); afterwards, each contract is judged extremely quickly (≈0.2 seconds per contract).

## 2 Background

**Solidity, bytecode, opcodes, basic blocks, control-flow graphs** A smart contract on the Ethereum blockchain is set of functions paired with some associated data, located at a specific address in the database. Most Ethereum smart contracts are written in a high-level language such as Solidity before being compiled to EVM bytecode and stored on the chain. EVM *bytecode* is represented by a very long number such as 0x6080604052348.... There is a simple injective relationship between valid hexadecimal sequences and a list of valid human-readable *opcodes* such as "PUSH1" (encoded as 0x60), "MSTORE" (0x52), and so forth. DLVA takes these hexadecimal bytecode sequences as input and disassembles them into opcode sequences; for readability we use opcodes hereafter. Ethereum's "Yellow Paper" defines the EVM as a variant of a stack machine with 150 distinct opcodes [79].

```
function withdraw() public {
    uint amount = balances[msg.sender];
    msg.sender.call{value: amount}("");
    balances[msg.sender] = 0; }
```

Figure 2: Sample representation of a program in Solidity

A *basic block* is an opcode instruction sequence without incoming or outgoing jumps, except at the beginning and end of the block. A *control-flow graph* (CFG) is a directed graph whose vertices are basic blocks and whose edges represent the execution flow among vertices. CFGs are more useful for analysis than linear representations of the code because they capture important semantic structures within the contract. In Figure 2 we give a sample Solidity smart contract; various standard lower-level representations are given in Appendix C.

**Smart contract vulnerabilities** Attacks on deployed smart contracts are commonplace [9]. DLVA hunts for the 29 vulnerability types summarized in Table 1. This includes many well-known vulnerabilities including "reentrancy-eth" (DAO, USD 50 million in losses [50, 66]) and "suicidal" (Parity, USD 280 million in losses [70]). Some well-known vulnerabilities in Table 1 are tagged with Smart contract Weakness Classification (SWC) numbers for ease of reference [58].

**EtherSolve, Slither** DLVA relies on two existing tools for smart contracts: EtherSolve [20] and Slither [25]. EtherSolve is used to disassemble EVM bytecode into opcode sequences and build the control flow graph. EtherSolve extracts the CFG using symbolic execution, resolving jumps symbolically.

The supervised training for our models uses Slither, a static analysis framework for Ethereum smart contracts, to label vulnerabilities. Slither recognizes the 29 vulnerabilities we list in Table 1, which is exactly why DLVA does too.

**Machine learning** DLVA is not "hardcoded" to understand the 29 vulnerabilities in Table 1. Instead, it is built on a series of carefully-chosen deep learning models (neural nets), which are trained on a large amount of data. Deep learning models come in two basic flavours: unsupervised and supervised. Unsupervised learning requires no input beyond the large dataset. Supervised learning requires an additional factor: each training input must be labeled by some external "supervisor."

**Universal Sentence Encoder (USE)** DLVA summarizes basic blocks with the USE [16]. USE encodes sentences into 512 dimensional vectors that encode sentence-level (in our context, basic block-level) similarity. USE has excellent performance for general text classification tasks and requires less training data than many competing machine learning techniques to build good predictive models. USE can be trained with or without supervision; we do not have labels for individual CFG nodes and so choose unsupervised training.

Table 1: 29 vulnerabilities in $\mathit{EthereumSC}_{large}$ (200+ times); ⋆ indicates 21 vulnerabilities in $\mathit{EthereumSC}_{small}$ (30+ times)

| | | Smart contract vulnerabilities | Large | Small |
|---|---|---|---|---|
| **High Severity** | ⋆ | **shadowing-state** (SWC-119): state variables with multiple definitions at contract and function level. | 3,602 | 52 |
| | ⋆ | **suicidal** (SWC-106): the `selfdestruct` instruction that is triggered by an arbitrary account. | 374 | 49 |
| | ⋆ | **uninitialized-state** (SWC-109): local storage variables are not initialized properly, and can point to unexpected storage locations in the contract. | 3,260 | 56 |
| | ⋆ | **arbitrary-send**: unprotected call to a function sending Ether to an arbitrary address. | 6,499 | 338 |
| | ⋆ | **controlled-array-length**: functions that allow direct assignment of an array's length. | 5,282 | 61 |
| | ⋆ | **controlled-delegatecall** (SWC-112): `delegatecall` or `callcode` instructions to external address. | 1,485 | 37 |
| | ⋆ | **reentrancy-eth** (SWC-107): usage of the fallback function to re-execute function again, before the state variable is changed (a.k.a. recursive call attack); reentrancies without Ether not reported. | 3,962 | 39 |
| | ⋆ | **unchecked-transfer**: the return value of an external transfer/transferFrom call is not checked. | 14,151 | 262 |
| **Medium Severity** | ⋆ | **erc20-interface**: incorrect return values for ERC20 functions. | 9,017 | 161 |
| | ⋆ | **incorrect-equality** (SWC-132): improper use of strict equality comparisons. | 8,604 | 95 |
| | ⋆ | **locked-ether**: contract with a payable function, but without withdrawal ability. | 12,164 | 398 |
| | | **mapping-deletion**: deleting a structure containing a mapping will not delete the mapping, and the remaining data may be used to breach the contract. | 235 | 0 |
| | | **shadowing-abstract**: state variables shadowed from abstract contracts. | 2,894 | 9 |
| | | **tautology**: expressions that are tautologies or contradictions. | 2,441 | 17 |
| | | **write-after-write**: variables that are written but never read and written again. | 467 | 6 |
| | ⋆ | **constant-function-asm**: functions declared as constant/pure/view using assembly code. | 4,019 | 49 |
| | | **constant-function-state**: calling to a constant/pure/view function that uses the `staticcall` opcode, which reverts in case of state modification, and breaking the contract execution. | 210 | 3 |
| | ⋆ | **divide-before-multiply**: imprecise arithmetic operations order; because division might truncate. | 14,529 | 176 |
| | ⋆ | **reentrancy-no-eth** (SWC-107): report reentrancies that don't involve Ether. | 14,982 | 130 |
| | | **tx-origin**: tx.origin-based protection for authorization can be abused by a malicious contract if a legitimate user interacts with the malicious contract. | 347 | 19 |
| | ⋆ | **unchecked-lowlevel**: return value of a low-level call is not checked. | 1,419 | 68 |
| | ⋆ | **unchecked-send**: return value of a send is not checked, so if the send fails, the Ether will be locked. | 593 | 68 |
| | ⋆ | **uninitialized-local** (SWC-109): uninitialized local variables; if Ether is sent to them, it will be lost. | 6,843 | 114 |
| | ⋆ | **unused-return** (SWC-104): return value of an external call is not stored in a local or state variable. | 11,222 | 2,427 |
| **Low Severity** | ⋆ | **incorrect-modifier**: modifiers that can return the default value, that can be misleading for the caller. | 1,273 | 171 |
| | ⋆ | **shadowing-builtin**: shadowing built-in symbols using variables, functions, modifiers, or events. | 1,536 | 35 |
| | ⋆ | **shadowing-local**: shadowing using local variables. | 26,259 | 174 |
| | | **variable-scope**: variable usage before declaration (*i.e.*, declared later or declared in another scope). | 1,484 | 27 |
| | | **void-cst**: calling a constructor that is not implemented. | 341 | 3 |

There are two USE variants: Deep Averaging Networks (DAN) [40] and the Transformer Architecture (TA) [76]. The TA produces more accurate models but required more computational resources to train than we had available. Fortunately, DAN's models are good enough to be very useful.

**Graph Neural Networks (GNNs)** DLVA's SC2V engine summarizes an entire CFG into an high-dimensional vector using a combination of GNN and traditional convolutional layers, which themselves use the 512-dimensional basic block summaries generated by USE as inputs. GNNs bring ideas from image processing to graph data [43]. The idea is that the interpretation of a pixel should be influenced by not only the contents of that pixel, but also the contents of neighboring pixels. To re-frame this idea into graphs, each pixel is a node, and edges connect neighboring pixels. In an image, each pixel-node has four immediate neighbors (except those nodes at the boundary of the image). A CFG is more complex since nodes are wired together in arbitrarily intricate ways, but the core idea is the same: aggregate the features of neighboring nodes with the features of the node itself into the summary.

**Feed Forward Network** The feedforward neural network (FFN) is one of the simpler kinds of neural network [60]. Information moves only forward (no cycles): from the input nodes, through any interior nodes, and lastly to the output nodes. During training, an FFN compares the outputs it generates with the oracle-dictated labels and then adjusts the edge weights to increase the accuracy in the next round.

**Evaluative metrics** Binary classification results are divided into *true positives* (TP), *true negatives* (TN), *false positives* (FP), and *false negatives* (FN). Derived metrics include *accuracy*; *true positive rate* (TPR), also known as recall, sensitivity, probability of detection, and hit rate; and *false positive rate* (FPR), also known as probability of false alarms and fall-out.

Although "accuracy" is important, it is not sufficient. Our data set is imbalanced: vulnerable smart contracts are scarce. Accordingly, a bogus model that simply labels all contracts as non-vulnerable will be surprisingly "accurate." Accordingly we also track TPR, which measures how often we catch vulnerable contracts; and FPR, which measures how often we issue false alarms. We formally define metrics in Appendix D.

# 3 Designing DLVA

Given a smart contract $c$ (expressed in bytecode) and vulnerability $v$ (from Table 1), our Deep Learning Vulnerability Analyzer's job is to predict label $c_v$, where $c_v = 1$ means that $c$ is vulnerable to $v$ and $c_v = 0$ means $c$ is secure from $v$.

Developing a tool that uses deep learning involves several steps. First, the overall architecture must be designed. Second, the resulting model must be trained on a suitable *training* data set. Third, substantial testing with a disjoint *validation* set is used to tune *hyperparameters*. These steps are the focus of §3. Evaluating the model on (disjoint) *testing* sets is in §4.

We further discuss the challenges in applying deep learning to smart contract vulnerability analysis in Appendix E.

**Overview** DLVA's design is in Figure 3. DLVA begins with the selection of a large training set, which is labeled for supervisory purposes as vulnerable or non-vulnerable for each attack vector. Next, a control-flow graph is extracted (§3.1).

The first neural net maps CFG Nodes to Vectors (N2V) using the Universal Sentence Encoder (USE), trained in unsupervised mode (§3.2). The second and third neural nets form the heart of DLVA. The Smart Contract to Vector (SC2V) engine maps smart contract into vectors; the Core Classifier (CC) classifies contracts as vulnerable or non-vulnerable by looking for 29 vulnerabilities. The design and training of these neural nets, including choices for hyperparameters, is in §3.3 and §3.4. Lastly, the Sibling Detector (SD) applies a simple heuristic to improve accuracy for "simple cases" (§3.5).

Once training has finished, analyzing a fresh contract proceeds as follows. First, bytecode is transformed to a CFG, and N2V summarizes its nodes into vectors. Next, SC2V uses those node summaries to summarize the entire CFG as a vector. This vector is given to the SD to see if it is close to a contract in the training set. If so, DLVA is done. If not, DLVA passes the vector to the CC, which renders its judgment.

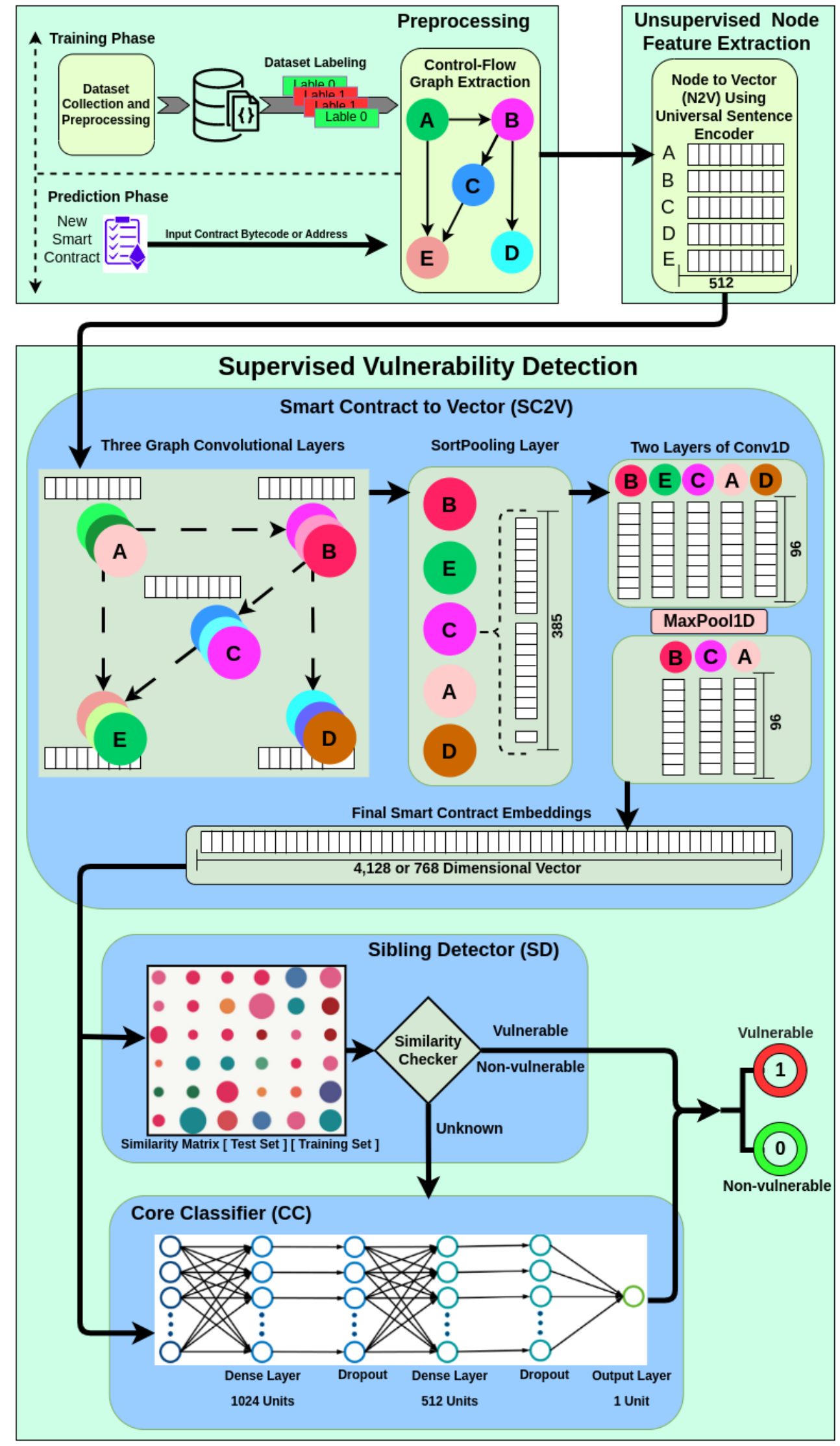

Figure 3: The Deep Learning Vulnerability Analyzer

## 3.1 Preprocessing

**Data Collection** We downloaded our Ethereum smart contract data set from Google BigQuery [33]. The dataset contains 51,913,308 contracts, but many are redundant: 99.3%, in fact. Removing redundant contracts leaves 368,438 distinct contracts, which we dub the *EthereumSC* data set.

**Data labeling** Two of our neural nets require labeled datasets to train. We chose Slither [25] (v0.8.0, committed on May 07, 2021, build 4b55839) to label because it covers a wide variety of vulnerabilities (74!), is more accurate than competitors for most vulnerabilities, and is relatively quick. Slither requires access to Solidity source code (rather than bytecode). Of the 368,438 contracts in our data set, only 120,365 (32.6%) had source available on Etherscan [24]. It took 13.6 days (using 1 core/16gb) for Slither to label them.

Quality training requires a reasonable number of positive examples, so we chose the 29 vulnerabilities that occurred at least 200 times for DLVA. Although some of the 29 vulnerabilities are more serious than others, it was troubling to discover that Slither considered only 37,574 contracts (31.3%) pristine from all 29 vulnerabilities. The remaining 82,609 vulnerable contracts had vulnerabilities distributed as shown in Table 1 (some contracts exhibit multiple vulnerabilities). Appendix F contains a graph of the frequency of the vulnerabilities.

DLVA must cope with messy realities, among them that Slither is impressive, but not foolproof. A manual inspection of 50 positive "reentrancy" vulnerabilities provided some evidence for a false positive rate of approximately 10% [25].

**Control-Flow Graph extraction** A Control-Flow Graph (CFG) makes a program's structure more apparent than a list of syntactic tokens does. We use EtherSolve [20] to disassemble a smart contract [79] and generate the associated opcode CFG. EtherSolve failed to create a CFG for 182 contracts (0.1% of the labelled data set), leaving us with 120,183 contracts suitable for training and testing. The average contract has 228 basic blocks, with 551 edges between them.

**Dividing our dataset** The deep learning techniques in DLVA work better if trained on contracts of similar size to the contract being analyzed, so we split the 120,183 distinct contracts with labels in *EthereumSC* into two datasets depending on length. The 7,017 contracts with fewer than 750 opcodes become the $EthereumSC_{small}$ data set, whereas the 113,166 contracts with between 750 and 10,000 opcodes become the $EthereumSC_{large}$ data sets. Both data sets are public [4, 3].

As is typical, we divide each data set into three disjoint subsets. The first 60% (in the order given by the Google BigQuery after filtering) we call the "training set," the next 20% is the "validation set," and the last 20% is the "test set."

## 3.2 Unsupervised Training: N2V

Sophisticated machine learning models typically work with numerical feature vectors rather than text. Our Node to Vector (N2V) component translates the opcode text within each CFG node (basic block) into such a feature vector to enable more sophisticated processing. We treat each basic block as a textual sentence of instructions (e.g. "PUSH1 0x80 PUSH1 0x40 MSTORE CALLVALUE..."). We then use the *Universal Sentence Encoder (USE)* [16] to transform these sentences into 512-dimensional vectors. We train USE by feeding it the ≈21.9 million basic blocks in our training & validation sets. We do not provide any expert rules or guidance.

As explained in §2, there are two variants of USE. The Transformer Architecture (TA) [76] yields more accurate models than Deep Averaging Networks (DAN) [40], at the cost of increased model complexity. We found the cost of TA too high: the 20-core/96-gb time-unlimited configuration ran out of memory, and the 12 hours available on the 24-core/180gb configuration were insufficient to finish training. DAN can be trained in linear time and was accurate enough for our purposes. To train DAN to summarize basic blocks took only 10.5 hours with a 12-core/16gb configuration.

```
Node        f0        f1  ...      f510      f511
-------------------------------------------------
0   -0.029561  0.022146  ...  -0.036117 -0.116207
12   0.009187  0.018610  ...   0.041887 -0.101014
...       ...       ...  ...        ...       ...
368 -0.015426  0.032837  ...  -0.062078  0.015451
371 -0.030309  0.054876  ...   0.009683 -0.023788
```

Figure 4: USE-generated vector embeddings

In Figure 4 we put USE/DAN's summary of the smart contract from Figure 2. Basic block nodes are labeled by the address of the first opcode in the block, and the f0 ... f511 give the corresponding 512-dimensional vector for the node.

## 3.3 Supervised Training: SC2V and CC

Our Smart Contract to Vector (SC2V) engine and Core Classifier (CC) form the heart of DLVA. As may be apparent from Figure 3, they have a relatively complex structure. Both are trained in supervised mode using the Slither-generated labels.

Although they are distinct components, SC2V and CC are trained together. Key idea: *by coupling their training we increase the accuracy of our predictive models.* Rather than having one universal SC2V model and 29 vulnerability-specific CC models, we actually have 29 SC2V/CC model pairs.

### 3.3.1 Smart Contract to Vector (SC2V)

Key idea: *treating programs as a graph rather than just a sequence of textual symbols increases the accuracy of our models*. SC2V maps smart contract CFGs to high-dimensional vectors. It takes as input the graph structure of the CFG (which we handle with Python's NetworkX library [36]), together with the USE-generated 512-dimensional vector embeddings for the associated basic blocks. We add self-loops to every node to increase the feedback in the neural net.

We use a modified *Graph Convolutional Network (GCN)* [43] combined with the SortPooling layer from the *Deep Graph Convolutional Neural Network (DGCNN)* [83] to analyze the complex structure of CFG graphs. We used three layers of GCN with 256, 128, and 1 neuron(s). The graph convolution aggregates a node's information with the information from neighboring nodes. The three layers propagate information to neighboring nodes (up to three "hops" away, and including the node itself due to self-loops), extracting local substructure and inferring a consistent node ordering.

We incorporated the SortPooling layer to sort the nodes using the third graph convolution (whose output is a single channel). After sorting the node summaries in ascending order by this channel, SortPooling selects the highest-valued 100 nodes, whose summaries are from the GCN layers, *i.e.* a 256 + 128 + 1 = 385-dimensional vector. We feed these sets of 385-dimensional vectors into a pair of traditional *Conv1D* convolutional layers, which further transform the 385-dimensional summaries into 96-dimensional vectors using rectified linear activation functions ("ReLU"). Between the Conv1D layers we use a *MaxPool1D* layer, which discards the half of the vectors with least magnitude. After the second Conv1D layer, we use a *Flatten* layer to produce the final vector representing the smart contract: contracts become 4,128-dimensional vectors.

#### 3.3.2 Core Classifier (CC)

As shown in Figure 3, the last neural net is a Feed Forward Network (FFN). The goal of the CC is to use the contract embeddings generated by SC2V to predict the label for arbitrary contracts. The structure of the FFN is three "Dense" layers with 1,024, 512, and 1 neuron(s) respectively. These layers use standard activation functions to activate said neurons: the first two layers use ReLU activation functions, whereas the final layer uses a sigmoid activation function. Between the layers we put standard "Dropout" filters with a 0.5 cutoff.

### 3.4 Selection of hyperparameters

Machine learning hyperparameters play a crucial role in model performance. Hyperparameters are set prior to training and affect the behavior of the learning algorithm. In our case we considered the following hyperparameters:

1. the number of graph convolutional layers (from {2, 3, 4}) and associated neuron sizes (from {128, 256});

2. aggregation methods (from {mean, sum, sort-top-k}), followed by {1, 2, 3} layers of Conv1D to reduce the size of the final embedding vector;

3. the number of Dense FFN layers (from {1, 2, 3}) with associated neuron sizes (from {256, 512, 1024}); and

4. activation functions {Hyperbolic Tangent, ReLU}.

In total we have $3 \times 2 \times 3 \times 3 \times 3 \times 3 \times 2 = 972$ possible hyperparameter settings. To reach the design given in Figure 3, we selected constant-function-asm in $EthereumSC_{large}$, trained all 972 possible models for that vulnerability, and selected the hyperparameters that performed best according to the validation set (disjoint from the training and test sets). We chose constant-function-asm because the number of positive examples were in the middle of the pack; the vulnerability is mostly only detected by Slither, thus minimizing the danger of biasing testing due to overfitting; and because we believed Slither's detector for this vulnerability was generally of high quality, with minimal false positives/negatives.

We used the same hyperparameters to train the other 28 models. In addition to giving us confidence that the models have not been overfitted, this implies that our architecture is relatively generic for smart contract vulnerability detection. DLVA does not rely on any manually designed expert rules or other human-generated hints. Accordingly, given suitable labeling oracles, training DLVA to recognize additional vulnerabilities is straightforward (we do this in §4.2 and §4.4/§B).

### 3.5 Sibling Detector (SD)

Given the smart contract embeddings generated by SC2V, we create a similarity matrix using Euclidean distance: $\sqrt{\sum_{i=1}^{N}(Q_i - P_i)^2}$, where $Q$ is a (previously unseen) contract embedding vector from the test set and $P$ is a contract embedding vector from the training set (with known label). The Sibling Detector labels $Q$ with the same label as the closest contract in the training set, as long as one exists within distance 0.1. Otherwise, the SD reports "unknown." SD starts with a distance of 0.0 and gradually increases it by 0.00001 until a contact is found or the maximum allowable distance of 0.1 is reached, whichever comes first. Sometimes $Q$ has multiple neighbors whose distances to $Q$ are within 0.00001. When this happens, the SD counts votes instead; if a strict majority are vulnerable, then SD reports $Q$ as "vulnerable."

### 3.6 Tweaking for smaller contracts

With the overall design settled, we make a few tweaks to better handle shorter contracts (under 750 opcodes). Since the there are many fewer distinct small contracts than large ones, we were only able to train 21 of the 29 vulnerability models on the $EthereumSC_{small}$ data set, despite lowering the minimum threshold to only 30 positive occurrences; we mark those 21 vulnerabilities in Table 1 with a $\star$.

We tweak SC2V's SortPooling layer to select the highest-valued 30 nodes (down from 100), which induces the Flatten layer to produce a 768-dimensional vector (down from 4,128). We use the same hyperparameters as for large contracts. The training set has fewer contracts so we turn off the SD.

### 3.7 Final details

**Engineering choices** We use Python's Keras framework to train our models. We train for 100 epochs (stopping early when *callbacks*=20). In each epoch, Keras feeds the networks the training set and adjusts their weights using the loss function *binary_crossentropy*. Keras uses the *Adam optimizer* with a categorical cross-entropy loss function to train more efficiently. We set the *learning_rate* to $5e-4$ and the *batch_size* to 512. We used BatchNormalization and Dropout layers to enhance the model's generalization and prevent overfitting.

Table 2: Datasets used for benchmarking DLVA

| Dataset | Contracts | Vul | Sz | Ground Truth |
|---|---|---|---|---|
| $EthereumSC_{large}$ [3] | 22,634 | 29 | L | Slither |
| $EthereumSC_{small}$ [4] | 1,381 | 21 | S | Slither |
| $Elysium_{benchmark}$ [2] | 900 (57) | 2 | S | Peer-reviewed |
| $Reentrancy_{benchmark}$ [6] | 473 (472) | 1 | S | P: Manual<br>N: 2 analyzers |
| $SolidiFI_{benchmark}$ [8] | 444 | 4 | L | P: Bug injection<br>N: 5 analyzers |
| $Zeus/eThor_{benchmark}$ [5] | 583 | 1 | S/L | Peer-reviewed |

**Training setup and time** A training machine has 96 GB of memory and a 20-core "Intel(R) Xeon(R) CPU E5-2650 v4 @ 2.20GHz" CPU. We used "CentOS Linux 7 (Core)," tensorflow 2.12.0 [1], tensorflow_hub 0.13.0, and Miniconda.

Since we had $972+28+21 = 1,021$ models to train, we used 10 training machines in parallel (200 cores, 960 GB). Total wall-clock training time was approximately four days.

## 4 Experiments and Evaluation

In §4 we evaluate the quality of DLVA from several complementary perspectives. In §4.1 we find and build benchmarks to help us understand DLVA's performance from a number of complementary perspectives. In §4.2 we analyze the performance of our neural nets vs. various ML alternatives, comparing our SC2V engine with four GNNs and our Core Classifier with ten machine learning alternatives. In §4.3 we measure how well DLVA can predict our oracle Slither. In §4.4 we compare DLVA with nine competitors. Lastly, in §4.5 we draw some conclusions from our experiments.

**Testing setup** Our test machine is a desktop with a 12-core 3.2 GHz Intel(R) Core(TM) i7-8700 and 16 GB of memory.

### 4.1 Designing benchmark datasets

It is challenging to pin down "ground truth" for tools that operate over large data sets. Most human-curated benchmarks contain fewer than 100 examples, and many of those are unrealistic, *e.g.* stripped to minimum size for pedagogical purposes. This is not how vulnerabilities occur in the real world.

Machine-curated benchmarks, such as $EthereumSC_{large}$ and $EthereumSC_{small}$ that we defined in §3.1, can contain large numbers of realistic contracts. However, it is hard to be totally confident about their labels. Tools capable of processing contracts at scale suffer from weaknesses that include: unsoundness, incompleteness, bugs, timeouts, and/or considering important classes of contracts to be out-of-scope.

The simple truth is that there are no existing benchmarks for Ethereum smart contract analysis tools that label large numbers of realistic contracts in a truly reliable way. We considered six benchmarks, summarized in Table 2, to help us evaluate DLVA from a variety of perspectives. "Contracts" indicates the number of contracts. Two benchmarks include contracts for which source code is unavailable; in this case the (parenthetical) gives the number that do have source code available. To help source code-based analyzers, in some cases we "lightly cleaned" the source code (*e.g.*, removing unicode from comments, moderately upgrading Solidity versions). "Vul" indicates the number of vulnerabilities; "Sz" whether the contracts are (mostly) Large (750–10,000 opcodes) or Small (less than 750); and the source of ground truth. All six datasets are disjoint from DLVA's training sets and are publicly available [3, 4, 8, 2, 6, 5]. We discussed $EthereumSC_{large}$ and $EthereumSC_{small}$ in §3.1; they are used in §4.2 and §4.3 to tell *how closely DLVA corresponds to Slither*. Three others are used to *evaluate DLVA's behaviour directly* in §4.4.

**$Elysium_{benchmark}$ [28]** This human-curated data set is by Torres *et al.*. $Elysium_{benchmark}$ combines the SmartBugs [26] and Horus [27] data sets for "Re-entrancy" (reentrancy-eth, 75 positive examples, 825 negative) and "Parity bug" (suicidal, 823 positive examples, 77 negative). $\text{Elysium}_{\text{benchmark}}$ contains many contracts that have been exploited in the real world. However, only 57 have available source (most suicided contracts are no longer available). We cleaned 2 contract sources. Most contracts are under 750 opcodes, a few 750-900.

**$Reentrancy_{benchmark}$** We sourced 53 contracts that exhibit the "Re-entrancy" (e.g. reentrancy-eth) vulnerability from the academic literature [41, 42, 59], reported attacks on GitHub, and various Ethereum blogs. We took well-reported vulnerabilities as positive ground truth. Almost all (52) had source code on Etherscan, and when so we manually confirmed the vulnerability. We cleaned 19 contract sources. We considered the 420 contracts that *both* Slither and Mythril labelled as safe from the 1,381 contracts in our $EthereumSC_{small}$ test set to be negative ground truth. All contracts are under 750 opcodes.

**$SolidiFI_{benchmark}$** To benchmark larger contracts, we used SolidiFI [32], a systematic method for bug injection that has been used in previous work to evaluate smart contract analysis tools [39, 82, 81, 56] to build the $SolidiFI_{benchmark}$.

Negative ground truth is established by the intersection of five static analyzers (Oyente, Mythril, Osiris, Smartcheck, and Slither). Positive ground truth is by injecting bugs from four different categories: Reentrancy (specifically, reentrancy-eth), Timestamp-Dependency, Overflow-Underflow, and tx.origin.

We generated 2,212 contracts, of which 80% were reserved for training/validation, with 20%—*i.e.*, 444 in total—available for testing, with each vulnerability occurring exactly 111 times. All contracts have source available, all of which we cleaned. All are over 750 opcodes. The contracts are complex but the injected vulnerabilities are simple; accordingly, performance may be better than for real-world vulnerable contracts. See Appendix G for more construction detail.

***Zeus/eThor$_{benchmark}$*** **[42, 62]** A reentrancy benchmark used to evaluate Zeus [42] and eThor [62]. Zeus's ground truth labels differ substantially from eThor's, making results hard to interpret and reinforcing the slippery nature of ground truth. Contract sizes are a mix of small and large; the subset we used have source. We discuss this dataset in Appendix H.

## 4.2 DLVA's neural nets vs. alternatives

**Node to Vector (N2V)** Appendix §I discusses other models/techniques we considered for N2V before settling on the Universal Sentence Encoder [16], including *fastText* [11], *word2vec* [51], Recurrent Neural Networks (RNNs) such as Long Short-Term Memory Networks (LSTMs) [38, 30], and Bidirectional Long Short-Term Memory (BiLSTMs) [63].

**Smart Contract to Vector (SC2V)** To evaluate SC2V we used *SolidiFI$_{benchmark}$*, since we consider its labels to be more reliable than *EthereumSC$_{large}$*. We used *SolidiFI$_{benchmark}$*'s training set for five state-of-the-art networks: a *Graph Convolutional Network* (GCN) [43], a *Gated Graph Sequence Neural Network* (GGC) [46], a *Graph Isomorphism Network* (GIN) [80], a *Deep Graph Convolutional Neural Network* (DGCNN) [83], and of course our own SC2V. For consistency, we trained all five competitors with DLVA's CC.

The results of our experiment are in Figure 5. (Appendix J contains four graphs isolating individual vulnerabilities.) We use the AUC "area under the receiver operating characteristic curve" metric, which measures the ability of the model to differentiate vulnerable from non-vulnerable cases, with higher scores better; AUC is explained further in Appendix D. SC2V has the highest score on Reentrancy and Overflow-Underflow, and ties with GCN for tx.origin; on Timestamp-Dependency, SC2V is a hair weaker than GCN. Averaged over all four vulnerabilities, SC2V leads with 99.0%, followed by GIN at 97.8%, GCN and GGC both at 97.5%, and finally DGCNN at 94.5%. Thus, SC2V beats the best competitor by 1.2% and the average competitor by 2.2%. SC2V performs better than competing models due to its more complex design: the convolutional layers of GCN and the SortPooling layer of DGCNN, followed by a pair of traditional convolutional layers.

**Core Classifier (CC)** In Figure 6 we benchmark CC against ten (well-trained) commonly used machine learning algorithms and one voting "meta-competitor." We trained all competitors on the *EthereumSC$_{large}$* training/validation sets and tested using the associated test set (*cf.* §3.1). We graph accuracy (higher is better), the True Positive Rate (higher is better); and the False Positive Rate (lower is better).

The ten established competitors have average accuracy of 68.7% (MLP's 71.1% is the highest). The voting meta-competitor reaches 71.6%. The CC's average accuracy of 80.0% crushes the competition by 11.3% and 8.4%.

In fact, DLVA's CC is more accurate than every other model, for every test.Moreover, the CC usually enjoys the highest/best TPR (or close), and the lowest/best FPR (or close). On a few tests the CC's TPR is uninspiring: uninitialized-state and write-after-write are the most challenging. Fortunately, on those difficult vulnerabilities, the CC's excellent FPR comes to the rescue. Conversely, the CC's FPR is uninspiring for constant-function-state; happily, it leads the pack on TPR.

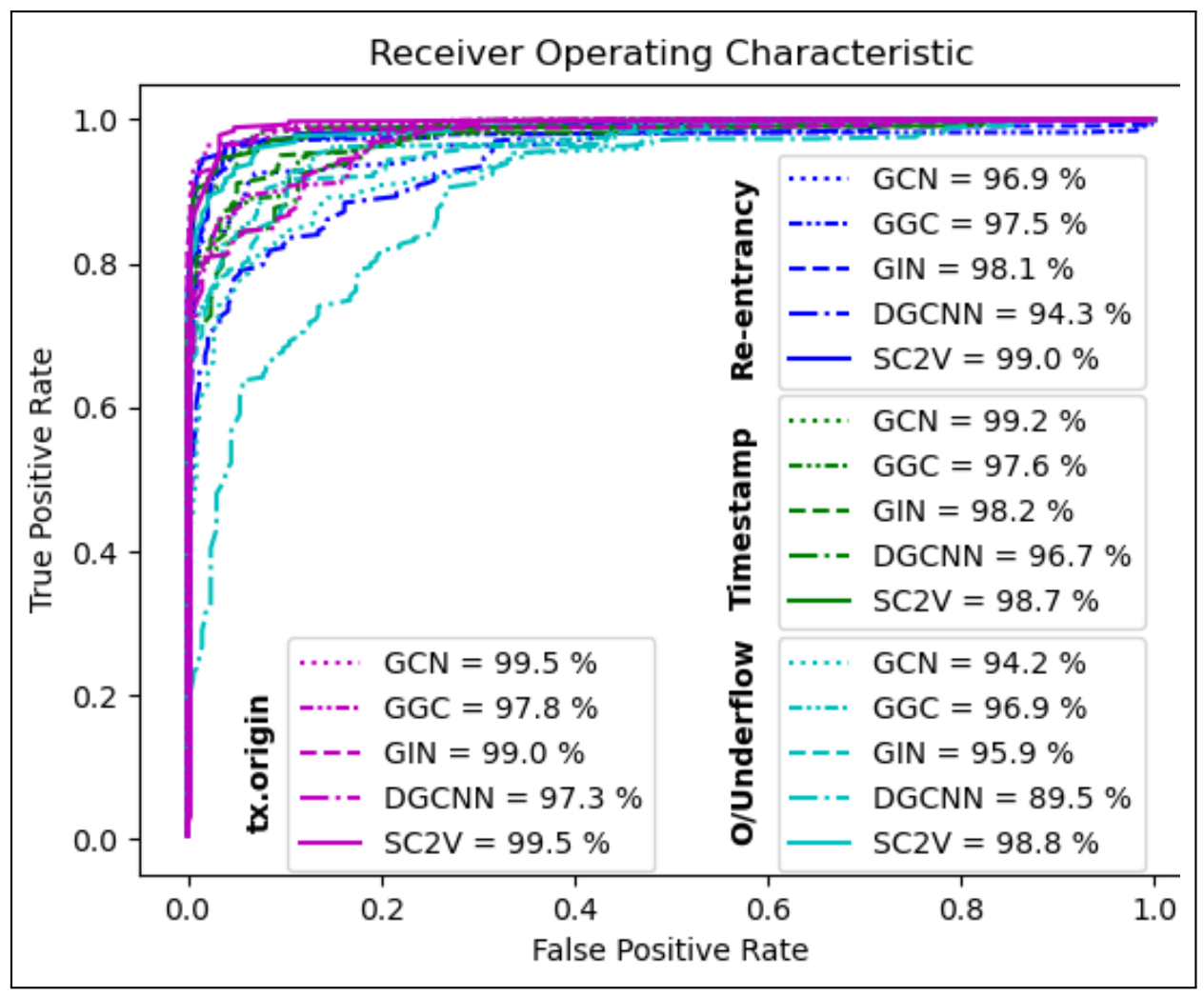

Figure 5: Evaluating SC2V vs. state-of-the-art GNNs

More information on this competition is in Appendix K.

## 4.3 Evaluating DLVA's models against Slither

Slither requires source code, whereas DLVA needs only bytecode. Only 32.6% of distinct contracts have source code available (§3.1); if DLVA accurately predicts Slither on those contracts, then it is probably accurately predicting how Slither would label the remaining 67.4%, were source code available.

Recall from §3.1 that we used Slither to label two different datasets: *EthereumSC$_{large}$* and *EthereumSC$_{small}$*. We used 60% of both data sets for training, and a further 20% for validation/tuning. The final 20% were not used in the development of DLVA and are thus suitable for evaluation (recall that the data sets contain distinct contracts, so no contract in the test set has been seen during training/validation).

### 4.3.1 *EthereumSC$_{large}$* results

Figure 7 summarizes the evaluation of 29 vulnerabilities with labels in *EthereumSC$_{large}$*. We measure three key statistics: on the left, accuracy (higher is better); in the middle, the True Positive Rate (higher is better); and at the right, the False Positive Rate (lower is better). The evaluative metrics were presented at the end of §2 and discussed further in Appendix D.

Each subgraph shows four distinct tasks, labeled CC-only for the Core Classifier on the entire test set, SD-easy for the Sibling Detector on 55.7% of the test set, CC-hard for the

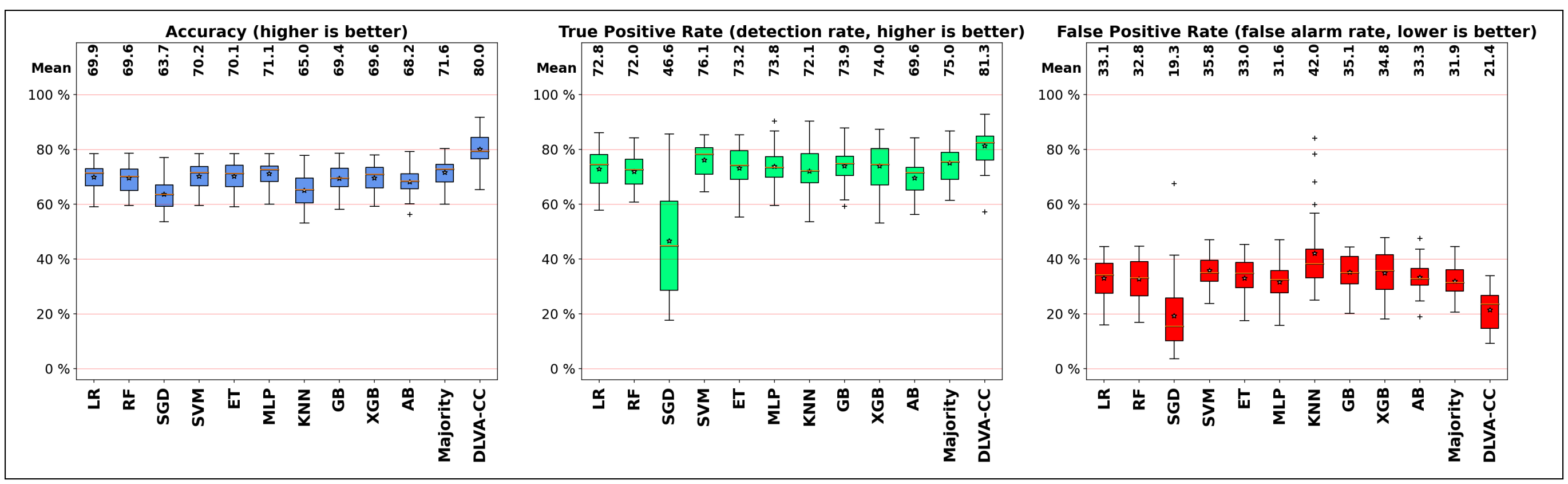

Figure 6: DLVA-CC vs. ten "off-the-shelf" ML classifiers and a majority voting strategy (⋆ is the average; + are outliers)

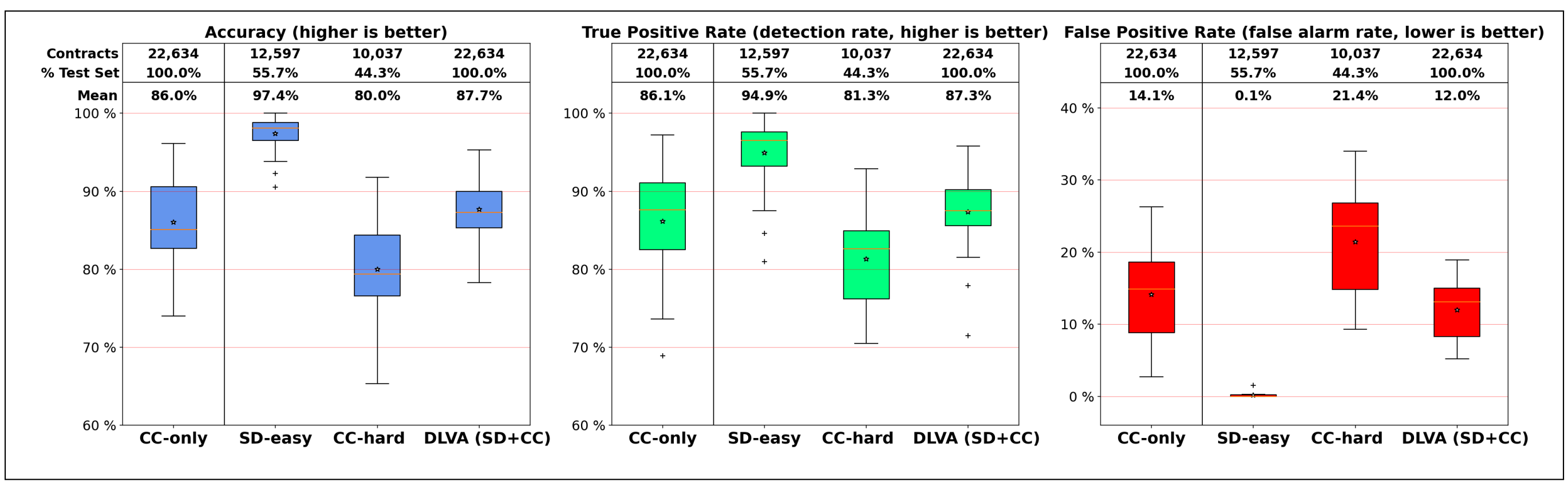

Figure 7: Deep Learning Vulnerability Analysis Tool Score Summary for 29 Vulnerabilities of *EthereumSC*$_{large}$ Dataset (The star symbol ⋆ represents the average, while the plus + represents outliers)

Core Classifier on the remaining 44.3%, and DLVA (SD+CC) for DLVA as a whole. The raw data is in Appendix L.

**Task CC-only: Core Classifier on entire dataset** We first measure the Core Classifier (CC) against the entire 22,634-contract test set. Average accuracy is 86.0%, TPR is 86.1%, and FPR is 14.1%. Appendix L contains the details for each vulnerability (Table Appx.9). Our next goal is to show that we can do better by incorporating our Sibling Detector.

**Task SD-easy: Sibling Detector** The Sibling Detector looks for smart contracts in the test set that are "very close" to contracts in the training set. Our distance threshold of 0.1 balances applicability and accuracy. At 0.1, a healthy 55.7% of the contracts in the test are close to a training set contract. To study accuracy, we ran the experiments reported in Appendix L (Table Appx.10). For the 12,597 test contracts (55.7%) within 0.1 distance of a training contract, SD achieved an accuracy of 97.4% with an FPR of only 0.1%. Accuracy was never lower than 90.5% and the FPR was never higher than 1.5%. The most challenging metric was TPR. Although average TPR was 94.9%, variance was higher. On the 5 most challenging vulnerabilities, TPR was 81.0%–89.5%.

**Task CC-hard: Core Classifier** We plan to use the CC only when the SD reports "unknown," *i.e.* the 10,037 contracts more than 0.1 away from any contract in the training set. The CC's job here is harder than in CC-only, the contracts are less similar to those seen during training. Despite this restriction, the CC had an average accuracy of 80.0% with an average FPR of 21.4% and TPR of 81.3%. The results for individual vulnerabilities are in Appendix L (Table Appx.11).

**Task SD+CC: DLVA as a whole** DLVA as a whole combines the SD and CC. If the SD can judge a contract, it does so. If not, DLVA uses the CC to make its best guess. DLVA has average accuracy of 87.7% with an associated FPR of only 12.0% and TPR of 87.3%. Appendix L reports per-vulnerability results (Table Appx.12).

Therefore, incorporating the SD into DLVA improves the statistics across the board: DLVA's accuracy goes up by 1.7%, its FPR goes down 2.1%, and its TPR goes up by 1.2%.

#### 4.3.2 *EthereumSC*$_{small}$ results

We also evaluated DLVA on 21 vulnerabilities with labels in *EthereumSC*$_{small}$. As shown in Appendix L (Table Appx.13),

Table 3: Comparison of DLVA vs. state-of-the-art tools; Input: (S: Source code, B:Bytecode, S/B$^-$: Source preferred, bytecode possible); Method: (SA:Static Analysis, FZ:Fuzzing, ML:Machine Learning, DL:Deep Learning); Vul: # of vulnerability detectors; Year: year of release of the used version; Cits: number of citations from Google Scholar on 29/5/2023.

| Analyzer | Input | Method | Vul | Year | Cits |
|---|---|---|---|---|---|
| **Oyente 0.2.7** [48] | S/B$^-$ | SA | 4 | 2017 | 1,996 |
| **Osiris** [72] | S/B$^-$ | SA | 5 | 2018 | 234 |
| **Mythril 0.21.20** [54] | S/B$^-$ | SA | 13 | 2019 | 127 |
| **SmartCheck 2.0** [71] | S | SA | 43 | 2019 | 513 |
| **SoliAudit** [47] | S | ML+FZ | 13 | 2019 | 76 |
| **eThor** [62] | B | SA | 1 | 2020 | 80 |
| **Slither 0.8.0** [25] | S | SA | 74 | 2021 | 292 |
| **ConFuzzius** [73] | S | SA+FZ | 10 | 2022 | 19 |
| **SAILFISH** [12] | S | SA | 2 | 2022 | 24 |
| **DLVA** | B | DL | 29 | 2023 | NA |

for such contracts DLVA has an average accuracy of 97.6% with a TPR of 95.4% and an associated FPR of only 2.3%.

#### 4.3.3 Overall fidelity to Slither

Averaging the separately-evaluated performance on both sizes of contract, DLVA has an overall average accuracy (to Slither) of 92.7%, a TPR of 91.4%, and a FPR of 7.2%.

### 4.4 DLVA vs. state-of-the-art tools

We selected the 9 competitors given in Table 3 to benchmark DLVA. We selected competitors based on a number of factors. We selected tools that require Source or those that can handle Bytecode; three competitors can handle both, but prefer source. Most competitors use some form of Static Analysis. We also included SoliAudit [47], the only publicly-available competitor tool using any form of Machine Learning. Two competitor tools use Fuzzing to augment their underlying analysis. DLVA is the first smart contract vulnerability analyzer using Deep Learning (neural nets). On average the competitors recognize 18 vulnerabilities, with significant variance. Table 3 also includes the year the version of the tool we used was released and a citation count for the underlying publication as a *very rough* measure of significance.

Benchmarking against multiple tools is inherently challenging. Many tools do not recognize the same vulnerabilities. *More seriously, even for the vulnerabilities that are recognized in common, the tools can define them differently*. Consider reentrancy, perhaps the most-studied vulnerability, and one recognized by all nine competitors. Recall from Table 1 that reentrancy actually comes in two flavours (reentrancy-eth and reentrancy-no-eth); this supported by the associated Solidity documentation [67]. However, only Slither (and, thus, DLVA) recognizes the -no-eth variety. If we include -no-eth examples in our benchmarks, other tools suffer many false negatives. Accordingly, -no-eth examples are not in our test sets.

To give another example, eThor [62] provides a formal definition for their notion of reentrancy ("single-entrancy"), and is the only competitor focused on soundness (*i.e.*, a 100% detection rate) above all else. However, single-entrancy considers unsafe some "litmus test" contracts that the SWC-107 description [55] labels safe[1]. Accordingly, eThor produces a lot of false positives when compared against a ground truth based on SWC-107. Moreover, eThor considers any contract containing a `DELEGATECALL` or `CALLCODE` opcode to be out of scope; in practice, this eliminates a *many* important examples.

**Summary of results** The high-level results of our competitor benchmarking was already given in Figure 1. Along the bottom we put the competitors, and in parenthesis the number of tests we include in the benchmark for that competitor (not every tool can handle every vulnerability).

We present five measures of performance. Overall, DLVA performed extremely well. The Completion Rate measures what percentage of contracts in our benchmarks terminate with a yes-or-no answer (rather than, *e.g.*, raising an exception, timing out, running out of memory). Most suffered from the occasional timeout or etc. Many of the source code analyzers were not able to analyze some contracts since the Solidity version was too old or new[2]. eThor refused to analyze many contracts due to `DELEGATECALL` or `CALLCODE` opcodes. Only DLVA and SmartCheck [71] answered every query.

Arguably the most important metrics are Accuracy, the True Positive Rate, and the False Positive Rate (see Appendix D for definitions). In Figure 1, we exclude any contract that failed to complete from these metrics (*i.e.*, we do not double count failures). eThor's focus on soundness paid off with a 100.0% TPR; Slither followed with 99.4%, and DLVA came in third with 98.7%. For FPR, SAILFISH boasts an impressive 0.1%, followed by DLVA at 0.6% and SmartCheck at 2.4%.

DLVA led the pack in accuracy at 99.7%, Slither came in second at 97.2%, and SmartCheck came in third at 93.2%. (Moreover, recall that DLVA judges bytecode whereas Slither and Smartcheck require source code!) DLVA's pack-leading accuracy is a result of, on the one hand, eThor's and Slither's higher/worse False Positive Rates; and on the other, SAILFISH's and SmartCheck's lower/worse True Positive Rates.

Average running time per contract in seconds is presented in Figure 1 on a logarithmic scale. DLVA is essentially an order of magnitude faster than Slither, its fastest competitor;

[1]For example, single-entrancy considers *both* the `simple_dao.sol` and `simple_dao_fixed.sol` litmus tests to be unsafe [61], whereas the SWC-107 description considers the first to be unsafe and the second to be safe [55].

[2]As mentioned in §4.1, we made a good-faith effort to lightly clean source code to help them, but in many cases it was not enough. We did exclude any contract for which source code was unavailable; Completion Rates would have been far worse for source-only competitors otherwise.

and three orders of magnitude faster than eThor, its slowest.

These benchmarks are presented in detail in Appendix B; the data underlying Figure 1 is in Tables 4, 5, and 6.

## 4.5 Discussion

Overall we are pleased with DLVA's performance as presented in this section: the machine learning models in DLVA are not trivial to best; DLVA is accurately predicting Slither's labels; and DLVA performs well compared to competitors. What remains is to highlight a few points and observations.

**Detecting vulnerabilities that caused heavy losses** Two smart contract losses loom large in the popular understanding: the DAO hack and the Parity bug. DLVA's performance on $Elysium_{benchmark}$ [2] showed that for real-world contracts with these vulnerabilities, DLVA's accuracy was 99.4%.

**70% of contracts-with-money are bytecode-only** Many existing tools—including Slither—require source code to analyze. In contrast, DLVA judges bytecode: essentially, extending Slither's "analysis style" from source- to bytecode.

We have identified approximately 12,000 contracts that hold at least 1 ETH each; the combined value is approximately 25,700,000 ETH (1 ETH is about 1,750 USD on June 12, 2023). Only 30% of these 12k contracts have source code available and are thus analyzable by Slither. In contrast, DLVA can judge all of them. We suggest that a user of a DLVA-flagged contract that lacks source code proceed with caution.

In our data set, we used DLVA to detect vulnerabilities in 248,073 contracts that were not labelled by Slither in §3.1 due to unavailability of source code on Etherscan, with total balance 540,928 ETH. DLVA flags about 6% of contracts for at least one high severity vulnerability.

**Speed matters for surveys and monitoring** DLVA is much faster than other tools. Running in "batch mode"—where all of the models are loaded into memory and then large numbers of contracts are analyzed—DLVA judges the average contract in 0.2 seconds. Slither takes 2 seconds per contract (10x slower), for the 32.6% of the contracts it can judge.

Some tools such as Mythril or eThor can analyze bytecode like DLVA. However, completion rates are much lower and the time required for analysis is 2-3 orders of magnitude greater. It is not practicable to scan large numbers of contracts for vulnerabilities with these tools, whether to survey the current state of the chain or to monitor new contracts in real time.

**Ongoing surveillance** Since DLVA is fast, accurate, and handles bytecode, we periodically run it over new contracts. When DLVA flags we "get a second opinion" from Slither/Oyente/Mythril, depending on whether source is available.

For example, DLVA recently flagged 95 contracts for the reentrancy vulnerability. These contracts jointly hold ≈85 ETH / 148k USD. Oyente and Slither confirmed the vulnerability for the 17 contracts that had source available, and Oyente and Mythril for the 78 that had only bytecode.

**Stability of vulnerability detection tools** A vulnerability classifier $X$ should give stable results: each time $X$ runs over a contract $c$ it should give the same answer. DLVA has this desirable property. We relabeled $EthereumSC_{large}$ with Slither and discovered that 1,328 labels changed from "vulnerable" to "non-vulnerable," and a further 172 labels changed from "non-vulnerable" to "vulnerable." Clearly Slither is not deterministic, perhaps due to timeouts or randomised algorithms. We estimate that Slither is mislabeling approximately 1.25% of contracts due to these issues. Clearly, DLVA's training algorithm is robust enough to cope with some mislabeling.

**Discovering label contradictions** We used the Sibling Detector to discover pairs of very similar contracts that Slither nonetheless labels differently. SD flagged them as potential label contradictions, reasoning that very similar contracts should have the same classification label. For example, for the "divide-before multiply" vulnerability, Slither labels the contract at address 0xaa3a2ae9 [18] as non-vulnerable and the contract at address 0xa8d8feeb [19] as vulnerable.

To resolve this apparent contradiction, we first asked DLVA's CC for its opinion (both considered non-vulnerable), and then manually examined the Solidity source code. Happily, DLVA's CC is right, whereas Slither's analysis of the contract at address 0xa8d8feeb is wrong. Further experiments with SD found 596 more "contradiction pairs." We manually reviewed 70 further pairs. For each pair reviewed, we found that both contracts had nearly identical solidity source code, differing only in initial values or whitespace/comments. We were pleased when the CC always assigned the same label to both contracts in a pair. After further manual inspection, we discovered that the CC was right 39 out of 71 times (55.0%).

This experiment indicates that machine learning techniques can help debug and improve static analyzers.

# 5 Related Work

The community has developed a variety of static analysis and dynamic analysis techniques to identify vulnerabilities in smart contracts. Static analyzers such as Oyente [48], Osiris [72], Mythril [54], SmartCheck [71], eThor [62], Slither [25], ConFuzzius [73], SAILFISH [12], Maian [57], Securify [74], and Manticore [53] rely on hand-crafted expert rules and manually engineered features. Although such tools are very impressive, and indeed we ourselves use Slither, this reliance on expert rules can make these tools difficult to maintain and update. We are unaware of any detection tool that detects all known vulnerabilities; or that is easily extendable for future bugs without human developers carefully crafting subtle expert rules and/or hardcoding additional features. Most smart contract vulnerability analyzers use symbolic execution to reason about all execution paths of a program. However, symbolic execution can suffer from "path explosion" when the size and complexity of the code increases, leading to sig-nificant time and space requirements. Practical limits on time and space can lead to difficulties analyzing smart contracts at scale. Moreover, empirical evaluation of 9 static analysis tools [23] classified 93% of contracts as vulnerable, thus indicating a considerable number of false positives. In addition, only a few vulnerabilities were detected simultaneously that got consensus from four or more tools.

Fuzzing is a dynamic analysis technique that has the advantage of scaling well to larger programs. Contractfuzzer [41], and Echidna [34] are two notable examples applied to smart contracts. Rather than relying on a fixed set of pre-defined bug oracles to detect vulnerabilities, fuzzing technique uses sophisticated grammar-based fuzzing campaigns based on a contract API to falsify user-defined predicates or Solidity assertions. However, generating meaningful inputs for fuzzing typically requires annotating the source code of a contract. Our benchmarking in §4.4 includes two hybrid tools that use fuzzing: ConFuzzius [73] and SoliAudit [47].

There is an ongoing trend of using machine learning (ML) for source and binary analysis of security concerns [49]. Recent surveys of machine learning techniques for source code analysis [65], malware analysis [75], and vulnerability detection [37] explored multiple categories for utilizing machine learning in code analysis. These categories include code representation [22], program synthesis [13], program repair [21], code clone detection [78], code completion [17], code summarization [77, 45], code review [44], code search [35, 14], and vulnerability analysis [31]. The breadth of this work shows that machine learning highlights the potential of these techniques for software development and security.

One of the closest pieces of related work to DLVA is Momeni et al. [52], which proposed a machine learning model to detect security vulnerabilities in smart contracts, achieving a lower miss rate and faster processing time than the Mythril and Slither static analyzers. Momeni et al.'s model is more handcrafted than DLVA, *e.g.* extracting 17 human-defined features from ASTs to measure the complexity of a small data set of source code. DLVA uses the Universal Sentence Encoder to extract features without human-provided hints.

Wesley et al. [69] adapted a long short-term memory (LSTM) neural network to analyze smart contract opcodes to learn vulnerabilities sequentially, which considerably improved accuracy and outperformed the symbolic analysis tool Maian. Compared with DLVA, Wesley et al.'s model learned from opcode sequences without considering the control flow of the smart contract, so it could not handle control-flow vulnerabilities. DLVA's choice to represent contracts as CFGs lets it understand more subtle vulnerabilities.

Sun et al. [68] added an attention mechanism to (non-graph) convolutional neural networks to analyze smart contract opcodes. They achieved a lower miss rate and faster processing time as compared to the Oyente and Mythril static analyzers. Liao et al. [47] developed SoliAudit, which combined machine learning and a dynamic fuzzer to strengthen the vulnerability detection capabilities. Liao et al. used word2vec to obtain a vector representation for each opcode and concatenated these vectors row-by-row to form the feature matrix. They did not consider the control-flow of the smart contract. In contrast, DLVA uses graph convolutional neural networks to extract contract embeddings, resulting in a more sophisticated understanding of program structure. Rather than combining with a fuzzer, we added our sibling detector SD.

SMARTEMBED [29] introduced the idea of clone detection for bug detection in Solidity code. SMARTEMBED used AST syntactical tokens to encode bug patterns into numerical vectors via techniques from word embeddings. Contracts are judged clones if they are Euclidian-close. The authors manually validated some reported bugs and showed that SMARTEMBED improved accuracy over SmartCheck. Our Sibling Detector was inspired by SMARTEMBED, although we work on CFGs rather than syntactic tokens. Moreover, SMARTEMBED was given predefined vulnerability matricies rather than learning from labeled data like DLVA.

Luca Massarelli et al. [49] investigated graph embedding networks to learn binary functions, proposing a deep neural network called structure2vec for graph embedding to measure the binary similarity of assembly code.

Some tool-specific details about the competitors benchmarked in §4.4 are given in Appendix B ("Discussion").

## 6 Conclusion

We have designed, trained, and benchmarked our novel Deep Learning Vulnerability Analyzer (DLVA). DLVA is an efficient, easy-to-use, and very fast tool for detecting vulnerabilities in Ethereum smart contracts. DLVA analyzes smart contract bytecode, so almost all smart contracts can be analyzed. DLVA transforms contract bytecode to an N-dimensional floating-point vector as a contract summary using our SC2V engine. This vector is given to DLVA's Sibling Detector to check whether it is Euclidian-close to contracts seen previously. If not, the vector is passed to DLVA's Core Classifier to predict the 29 vulnerabilities learned during training.

DLVA has a generic design, rather than one customized for each vulnerability. Accordingly, given bytecodes and suitable labeling oracles, training DLVA to recognize future smart contract vulnerabilities should be straightforward without the need for expert rules and/or hardcoding additional features.

As shown in Figure 1, DLVA outperformed nine state-of-the-art alternatives. DLVA leads the pack with a 100% Completion Rate, 99.7% Accuracy, and 0.2 second average contract analysis time, *i.e.*, 10-1,000x faster than competitors. DLVA's True Positive Rate of 98.7% and False Positive Rate of 0.6% are highly competitive too (#3 and #2, respectively).

**Acknowledgements** We thank the CRYSTAL Centre (NUS) and Joxan Jaffar for financial support; and the anonymous reviewers and shepherd for their many suggestions.

## Availability

DLVA is available for download from **https://bit.ly/DLVA-Tool** (see "README.md"). The data sets we use in this paper are available as well [3, 4, 8, 2, 6].

## A Appendix Overview

Appendix B, in both this paper and the conference version [7], details the comparison in §4.4. The other appendices (§C–L) are unique to this extended version and organized as follows. Appendices §C and §D extend §2 to detail smart contract code representations and the evaluation metrics we use. Appendices §E and §F extend §3 to discuss challenges in applying deep learning to smart contract vulnerability analysis and to graph the frequencies of vulnerabilities in the EthereumSC dataset. Appendices §G and §H extend §4.1 to detail the construction of the SolidiFI benchmark and to discuss the $Zeus/eThor_{benchmark}$. Appendices §I, §J, and §K extend §4.2 to discuss node feature extraction (N2V), to separate the individual SC2V comparisons overlaid in Figure 5, and to detail the comparison of DLVA-CC with 10 ML classifiers and the voting meta-competitor from in Figure 6. Appendix §L extends §4.3 to include the numerical tables behind Figure 7.

## B Details of the Competitor Benchmark in §4.4

Tables 4, 5, and 6 contain the data behind Figure 1. We benchmark with $Elysium_{benchmark}$ [2], $Reentrancy_{benchmark}$ [6], and $SolidiFI_{benchmark}$ [8] since we have high confidence in their labelling of ground truth (§4.1). We discuss the $Zeus/eThor_{benchmark}$ [5] and its challenges in Appendix H.

Table 4 presents the results of the five bytecode analyzers on $Elysium_{benchmark}$ [2], $Reentrancy_{benchmark}$ [6]. $Elysium_{benchmark}$ [2] contains contracts with two vulnerabilities: REentrancy (with 75 Genuine/ground Positives and 825 Genuine/ground Negatives) and Parity Bug (with 823 GP and 77 GN). $Reentrancy_{benchmark}$ only contains reentrancies (with 53 GP and 420 GN). For each benchmark, we document five statistics for each tool. 'Exp" gives the number of contracts for which analysis failed to complete (*e.g.*, timeouts). False Negatives gives the number of ground positives that were incorrectly labeled as negative; conversely, False Positives gives the number of ground negatives that were incorrectly labeled as positive. The Σum of Failures is just Exp + FN + FP, and the Average Failure is the number of failures as compared to the size of the test set (averaging the failure rate for RE and PB for the two tools that can handle both). Table 5 gives the same data for source code analyzers. Only 6.6% of contracts in $Elysium_{benchmark}$ have available source (from 900 contracts to 59) since suicided PB contracts no longer have source code on Etherscan. We considered marking the 841 missing contracts as "Exp" to emphasize the importance analyzing bytecode, but ultimately decided to simply exclude them. Table 6 gives the associated data for $SolidiFI_{benchmark}$.

**Training** Slither does not recognize Overflow/Underflow and $EthereumSC_{large}$ had too few occurrences of Timestamp-Dependency to be included in the 29 vulnerabilities in Table 1. We retrained DLVA to handle these vulnerabilities with the training and validation portions of the SolidiFI dataset.

**Discussion** What follows is a brief explanation of individual competitors and our understanding of their performance.

Oyente is a symbolic execution tool, it sets limits on loop depth (10) and path depth (50) to decrease space explosion. However, this leads to a significant increase in false negatives.

Osiris builds on Oyente, using the same limits. It combines symbolic execution and taint analysis to enhance the vulnerability detection, especially for small contracts.

Mythril employs concolic analysis, taint analysis, and control flow checking of EVM bytecode to effectively narrow down the search space. However, Mythril experiences 2.4% exceptions within the analyzed contracts and demonstrates slower performance when compared to competing tools.

Smartcheck utilizes an intermediate representation (IR) generated from the source code and then scans this representation using XPath patterns to detect bugs. However, the XPath patterns employed by Smartcheck are sensitive to even slight variations in the syntax of bug snippets. As a result, Smartcheck reduces the occurrence of false positives to 0, but exhibits a notable number of false negatives.

SoliAudit, a machine learning- and fuzzing-based vulnerability analyzer, may have missed injected bugs that differed significantly from the patterns it learned during training.

eThor prioritizes soundness, with a "single-entrancy" definition that conservatively approximates true reentrancies. However, this approach results in a much higher number of false positives and a significant number of exceptions due to the restrictions imposed by their definition.

Slither exhibits exceptional performance in bug detection, second only to DLVA in overall accuracy, effectively capturing bugs that fall within its defined scope and definitions.

ConFuzzius uses a hybrid of symbolic execution and fuzzing with a dynamic data dependency analysis to identify vulnerabilities. ConFuzzius's Completion Rate of only 89.9% is due in part to not supporting solc version < 0.4.11.

Sailfish, built on Slither, shares its speed. Compared to Slither, Sailfish has only one false positive but more false negatives, perhaps due to a difference between Sailfish's definition for reentrancy and real-world bugs. Sailfish's Completion Rate of 87.8% is due to building on an old version of Silther that is not compatible with new versions of Solidity.

Table 4: Small contracts, bytecode analyzers; Exp: Exceptions; Vulnerability: {RE:Reentrancy, PB:Parity Bug}; GP: Ground Positives; GN: Ground Negatives; FN: False Negatives; FP: False Positives; ΣF: Sum of Failures

| Analyzer | **Benchmark Data Sets** (entire benchmark) | | | | | | | | | | | | |
|---|---|---|---|---|---|---|---|---|---|---|---|---|---|
| | **Elysium** [2] | | | | | | | | **Reentrancy** [6] | | | | |
| | | **RE** 75 GP + 825 GN | | | **PB** 823 GP + 77 GN | | | Average Failure | | **RE** 53 GP + 420 GN | | | Average Failure |
| | Exp | FN | FP | ΣF | FN | FP | ΣF | | Exp | FN | FP | ΣF | |
| **Oyente** [48] | 1 | 28 | 0 | 29 | - | - | - | **3.2%** | 12 | 27 | 0 | 39 | **8.2%** |
| **Osiris** [72] | 0 | 6 | 0 | 6 | - | - | - | **0.7%** | 12 | 2 | 3 | 17 | **3.6%** |
| **Mythril** [54] | 37 | 3 | 0 | 40 | 819 | 0 | 856 | **49.8%** | 39 | 0 | 3 | 42 | **8.9%** |
| **eThor** [62] | 830 | 0 | 1 | 831 | - | - | - | **92.3%** | 287 | 0 | 130 | 417 | **88.1%** |
| **DLVA** | 0 | 1 | 4 | 5 | 1 | 3 | 4 | **0.5%** | 0 | 3 | 0 | 3 | **0.6%** |

Table 5: Small contracts, source code analyzers; Exp: Exceptions; Vulnerability: {RE:Reentrancy, PB:Parity Bug}; GP: Ground Positives; GN: Ground Negatives; FN: False Negatives; FP: False Positives; ΣF: Sum of Failures

| Analyzer | **Benchmark Data Sets** (subset of benchmark with available source code) | | | | | | | | | | | | |
|---|---|---|---|---|---|---|---|---|---|---|---|---|---|
| | **Elysium** [2] | | | | | | | | **Reentrancy** [6] | | | | |
| | | **RE** 52 GP + 7 GN | | | **PB** 7 GP + 52 GN | | | Average Failure | | **RE** 52 GP + 420 GN | | | Average Failure |
| | Exp | FN | FP | ΣF | FN | FP | ΣF | | Exp | FN | FP | ΣF | |
| **SmartCheck** [71] | 0 | 5 | 1 | 6 | - | - | - | **10.2%** | 0 | 0 | 1 | 1 | **0.2%** |
| **SoliAudit** [47] | 7 | 11 | 1 | 18 | 0 | 1 | 8 | **22.0%** | 88 | 10 | 5 | 103 | **21.8%** |
| **Slither** [25] | 1 | 1 | 6 | 8 | 0 | 0 | 1 | **7.6%** | 21 | 1 | 0 | 22 | **4.7%** |
| **ConFuzzius** [73] | 7 | 7 | 1 | 15 | 0 | 0 | 7 | **18.6%** | 111 | 2 | 0 | 113 | **23.9%** |
| **SAILFISH** [12] | 6 | 16 | 0 | 22 | - | - | - | **37.3%** | 125 | 24 | 1 | 150 | **31.8%** |

Table 6: Large contracts; Exp: Exceptions; Vulnerability: (RE:Reentrancy, TS:Timestamp-Dependency, OU:Over/Underflow, TX:tx.origin); GP: Ground Positives; GN: Ground Negatives; FN: False Negatives; FP: False Positives; ΣF: Sum of Failures

| Analyzer | **SolidiFI** [8] (entire benchmark) | | | | | | | | | | | | | |
|---|---|---|---|---|---|---|---|---|---|---|---|---|---|---|
| | | **RE** 111 GP + 333 GN | | | **TS** 111 GP + 333 GN | | | **OU** 111 GP + 333 GN | | | **TX** 111 GP + 333 GN | | | Average Failure |
| | Exp | FN | FP | ΣF | FN | FP | ΣF | FN | FP | ΣF | FN | FP | ΣF | |
| **Oyente** [48] | 0 | 0 | 0 | 0 | 110 | 19 | 129 | 12 | 267 | 279 | - | - | - | **30.6%** |
| **Osiris** [72] | 0 | 10 | 0 | 10 | 111 | 17 | 128 | 20 | 240 | 260 | - | - | - | **29.9%** |
| **Mythril** [54] | 0 | 68 | 23 | 91 | 43 | 15 | 58 | 81 | 85 | 166 | 23 | 7 | 30 | **19.4%** |
| **SmartCheck** [71] | 0 | 0 | 0 | 0 | 52 | 0 | 52 | 83 | 0 | 83 | 0 | 0 | 0 | **7.6%** |
| **SoliAudit** [47] | 0 | 111 | 0 | 111 | 1 | 21 | 22 | 13 | 282 | 295 | 0 | 7 | 7 | **24.5%** |
| **eThor** [62] | 194 | 0 | 135 | 329 | - | - | - | - | - | - | - | - | - | **74.1%** |
| **Slither** [25] | 0 | 0 | 0 | 0 | 0 | 20 | 20 | - | - | - | 0 | 0 | 0 | **1.5%** |
| **ConFuzzius** [73] | 7 | 54 | 2 | 63 | - | - | - | 64 | 58 | 129 | - | - | - | **21.6%** |
| **SAILFISH** [12] | 0 | 0 | 0 | 0 | - | - | - | - | - | - | - | - | - | **0.0%** |
| **DLVA** | 0 | 0 | 0 | 0 | 0 | 0 | 0 | 2 | 0 | 2 | 0 | 0 | 0 | **0.1%** |

# Smart Learning to Find Dumb Contracts
## (Extended Appendix)

Tamer Abdelaziz[†]
tamer@comp.nus.edu.sg
(†) National University of Singapore
Singapore

Aquinas Hobor[‡,†]
a.hobor@ucl.ac.uk
(‡) University College London
London, United Kingdom

**References in the Extended Appendix** This extended appendix has its own list of references at the end. Reference numbers do not refer to the reference list in the main paper, but instead to the reference list in the extended appendix.

## C Contract Representations (extending §2)

**Contract Code:** Smart contracts are initially written in a high-level programming languages such as Solidity, Vyper or Serpent. The source code contains the logic and functionality of the contract and is written by developers. It includes functions, variables, modifiers, events, and other elements that define the behavior of the contract, as shown in Figure Appx.8. This code is the full contract partially excerpted in Figure 2.

The Solidity compiler, called "solc", is used to compile the source code into bytecode that can be executed by the EVM. The bytecode is a sequence of machine instructions that the EVM can understand and execute. It is stored on the blockchain when the contract is deployed and is immutable once deployed, as shown in Figure Appx.9. The opcodes serve as the underlying instructions that enable the EVM to interpret and execute the specific operations defined within a smart contract bytecode. They provide the essential means for carrying out various tasks and implementing the desired functionalities within the Ethereum ecosystem, as shown in Figure Appx.10. There is a simple injective relationship between valid hexadecimal bytecode sequences and a list of valid human-readable opcodes such as "PUSH1" (encoded as 0x60), "MSTORE" (0x52), and so forth.

```
pragma solidity >=0.7.0 <0.9.0;
contract Mybank {
  mapping(address => uint) private balances;
  function withdraw() public {
      uint amount = balances[msg.sender];
      msg.sender.call{value: amount}("");
      balances[msg.sender] = 0;
      }
  }
```

Figure Appx.8: Full smart contract for Figures 2 and 4

```
0x6080604052348015610010576000080fd5b5060043610
002b5760003560e01c80633ccfd60b14610030575b6000
...
735822122083886215b809901e5316a29fb7b300e0026d
b9a3a9fb143816c40ffe5b89d13464736f6c6343000807
```

Figure Appx.9: Smart Contract Bytecode.

```
PUSH1 0x80 PUSH1 0x40 MSTORE CALLVALUE DUP1
ISZERO PUSH2 0x10 JUMPI PUSH1 0x0 DUP1 REVERT
...
MUL PUSH14 0xB9A3A9FB143816C40FFE5B89D13 PUSH5
0x736F6C6343 STOP ADDMOD SMOD STOP CALLER
```

Figure Appx.10: Smart Contract Opcodes.

**Control Flow Graph:** The Ethereum bytecode control flow graph represents the flow of execution and control flow within a smart contract in the form of a graph. It involves analyzing the bytecode to identify the sequence of opcodes and their associated jump instructions to provide a visual representation of how the program's execution moves from one instruction to another based on different conditions and control statements, as shown in Figure Appx.11. Constructing a precise control flow graph for Ethereum bytecode can be challenging since EVM bytecode jump destinations are read from the stack and hence can be subject to prior computations.

The node identifiers in Figure 4 do not match up perfectly to the CFG pictured in Figure Appx.11, which gave only a partial contract. Figure 4 summarizes the entire contract, including the boilerplate from Figure Appx.8 hidden in Figure 2.

We use CFGs as the basis of our SC2V engine because [23] indicated that their more-linear approach limited their ability to handle vulnerabilities that depended on control flow.

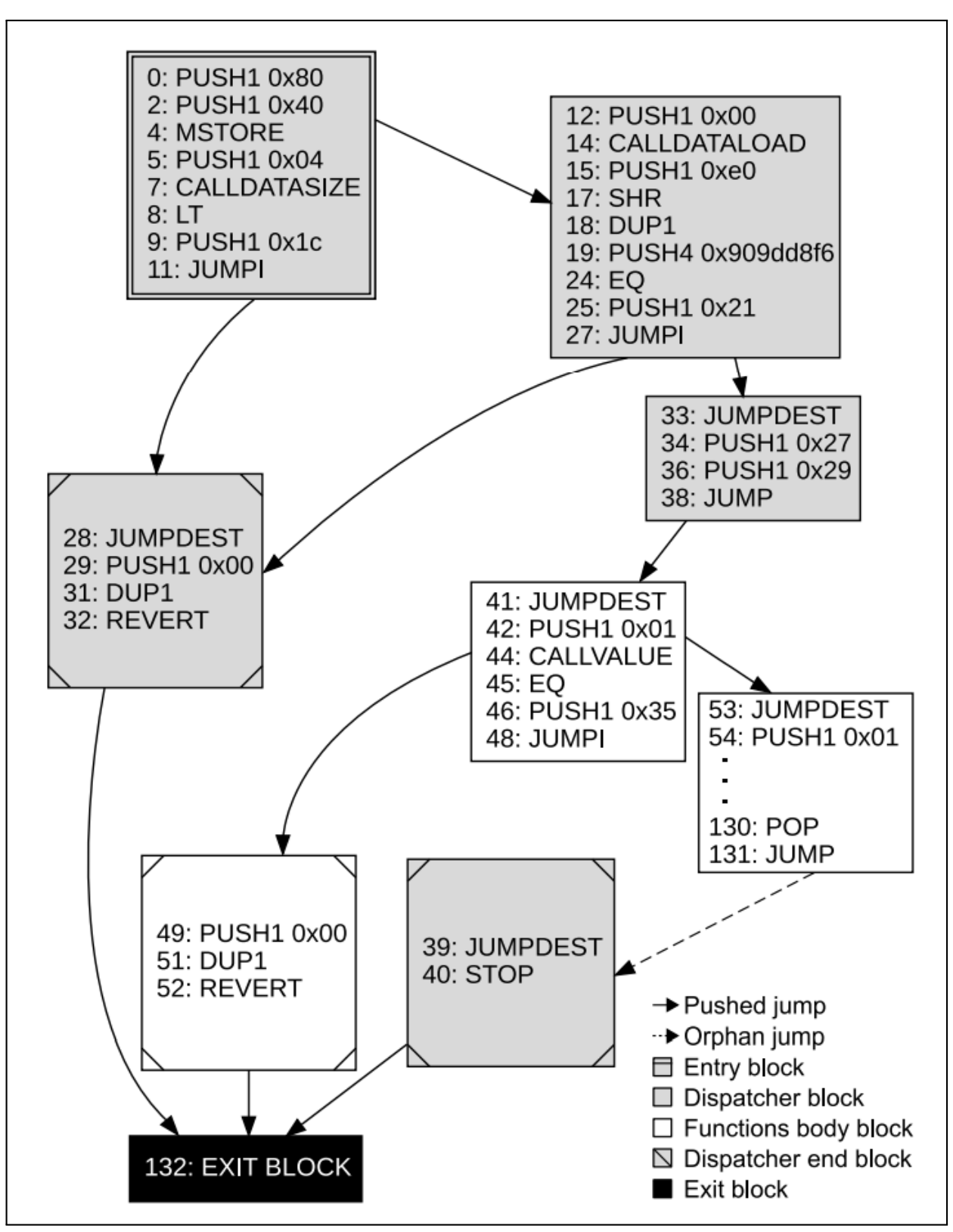

Figure Appx.11: Smart Contract Control Flow Graph (adapted from [3]).

**Advantages and Disadvantages of Bytecode Analysis** Analyzing smart contract bytecode offers several distinct advantages. First, bytecode analysis avoids any errors introduced during compilation: the analyzed code is really what runs on the chain. Second, bytecode is immutable once deployed on the blockchain, ensuring that the analyzed code remains constant indefinitely. Sometimes—either due to mistake or malice—the published source code for a contract is not accurate. Third, many contracts avoid publishing their source code, *e.g.* to partially shield sensitive business logic. Fourth, many of the source code analyzers we investigated in §4.4 had major problems with versioning (Solidity versions that were too old or too new) and other in-theory irrelevant issues such as Unicode characters in code comments. We cleaned a lot of source code in §4.1 to try to help the source code analyzers get around these kinds of problems; bytecode analyzers never worry about them.

On the other hand, sometimes source code can help. For example, the uninitialized-state and -local vulnerabilities (SWC-109) are obvious in source, but must be inferred indirectly in bytecode, *i.e.*, when all variables "happen" to be initialized to 0. DLVA's relatively poor performance on these two vulnerabilities (*cf.* Table Appx.12) is exactly due to some information at source code being "compiled away" in bytecode.

# D Evaluation Metrics (extending §2)

The binary classification has four possible outcomes: true positives (TP), true negatives (TN), false positives (FP), and false negatives (FN). We adopted the evaluation metrics of accuracy, true positive rate (TPR), true negative rate (TNR), false positive rate (FPR), false negative rate (FNR), and area under the curve (AUC) scores on the test dataset. We could not use accuracy in (1) as the only metric for evaluation because our dataset is imbalanced (the total number of vulnerable smart contracts are scarce compared to the non-vulnerable), as the model could easily achieve high accuracy by labeling all samples as the majority class (non-vulnerable class) and neglect the minority (vulnerable class). Accuracy works best if false positives and false negatives have similar cost. In our problem false negative has dire consequences more than false positive.

$$\text{Accuracy} = \frac{\text{True Positives} + \text{True Negatives}}{\text{Total Tested}} \tag{1}$$

True positive rate in (2) (also known as recall, sensitivity, probability of detection, hit rate) and false positive rate in (3) (also known as probability of false alarm, fall-out) are the other two metrics frequently adopted to evaluate a binary classification model.

$$\text{True Positive Rate} = \frac{\text{True Positives}}{\text{True Positives + False Negatives}} \tag{2}$$

$$\text{False Positive Rate} = \frac{\text{False Positives}}{\text{True Negatives + False Positives}} \tag{3}$$

The value of TPR measures the ability not to miss any vulnerable contract (how many of the actual positives our model is able to capture through labeling it as positive and it is true positive), while the value of FPR measures the ability of the model to reduce false alarms. These metrics help to measure the detection rate of vulnerable contracts, and calculate the false alarms of mislabeling an innocent contract as vulnerable.

The receiver operating characteristic (ROC) curve is frequently used for evaluating the performance of binary classification algorithms. ROC is produced by calculating and plotting the true positive rate against the false positive rate for a single classifier at a variety of thresholds. AUC stands for area under the ROC curve and represents the degree or measure of separability. AUC tells how much the model is capable of distinguishing between non-vulnerable and vulnerable classes[3].

# E Challenges and Opportunities For Machine Learning in Smart Contract Vulnerability Analysis (extending §3)

Smart contracts pose unique challenges compared to general programming languages. Most source code is unavailable.

[3] AUC visualization: https://arize.com/blog/what-is-auc/

Vulnerabilities can cause significant financial losses so the incentives to attack are high.

The primary competitors for DLVA are based on static analysis. Developing static analyzers requires devising subtle expert rules that characterize the vulnerabilities of interest. This is not easy. Four of the nine competitors in Table 3 handle fewer than 5 vulnerabilities; another three handle 10-13. Only SmartCheck, with 43; and Slither, with 74; handle more vulnerabilities than DLVA (29+). In contrast, training a new model for DLVA requires only a suitably large dataset: in §3.3 we train 29 models using $EthereumSC_{large}$, in §3.6 we train 21 more models using $EthereumSC_{small}$, and in §4.4/§B we train a final two models using $SolidiFI_{training}$. Deep learning avoids the difficult work of developing subtle expert rules.

Of course, ML models bring their own challenges. A major one is the need to find or develop ethically appropriate high-quality labelled datasets: *i.e.*, with generally trustworthy labels, a reasonable supply of both positive and negative occurrences, and limited concerns regarding privacy and bias. Our use of unsupervised training for N2V in §3.2 does lower this cost somewhat. A second challenge is the major computational requirements for training, although a balancing factor is the relatively low requirements for running afterwards (*e.g.*, eThor requires no time to train, but does require 1,000x time to run). A third challenge is the necessity of an "experimental science" approach to find a "good" model (*e.g.* by tuning hyperparameters). Lastly, using deep learning to identify zero-day vulnerabilities seems especially challenging, as the nature of the problem seems to exclude labelled data for training.

## F Vulnerability Frequencies in the EthereumSC Dataset (extending §3)

Figure Appx.12 visually presents the vulnerability frequencies within the EthereumSC dataset, showing the distribution and occurrence of each vulnerability type with a bar graph. Each bar corresponds to a specific vulnerability type, while the height of the bars signifies the frequency or occurrence of that particular vulnerability type within the dataset.

## G Construction of the SolidiFI benchmark (extending §4.1)

***$SolidiFI_{benchmark}$*** We develop $SolidiFI_{benchmark}$ to evaluate performance on larger contracts using SolidiFI [6], a systematic method for bug injection that has been used in previous work to evaluate smart contract analysis tools [12, 27, 25, 18]. SolidiFI is used to inject security bugs into Solidity smart contracts. In [6] SolidiFI is employed to evaluate all ten analysis tools for Ethereum contracts, focusing on false negatives and false positives. SolidiFI, the tool, formulates distinct code snippets representative of exploitable vulnerabilities corresponding to each bug type.

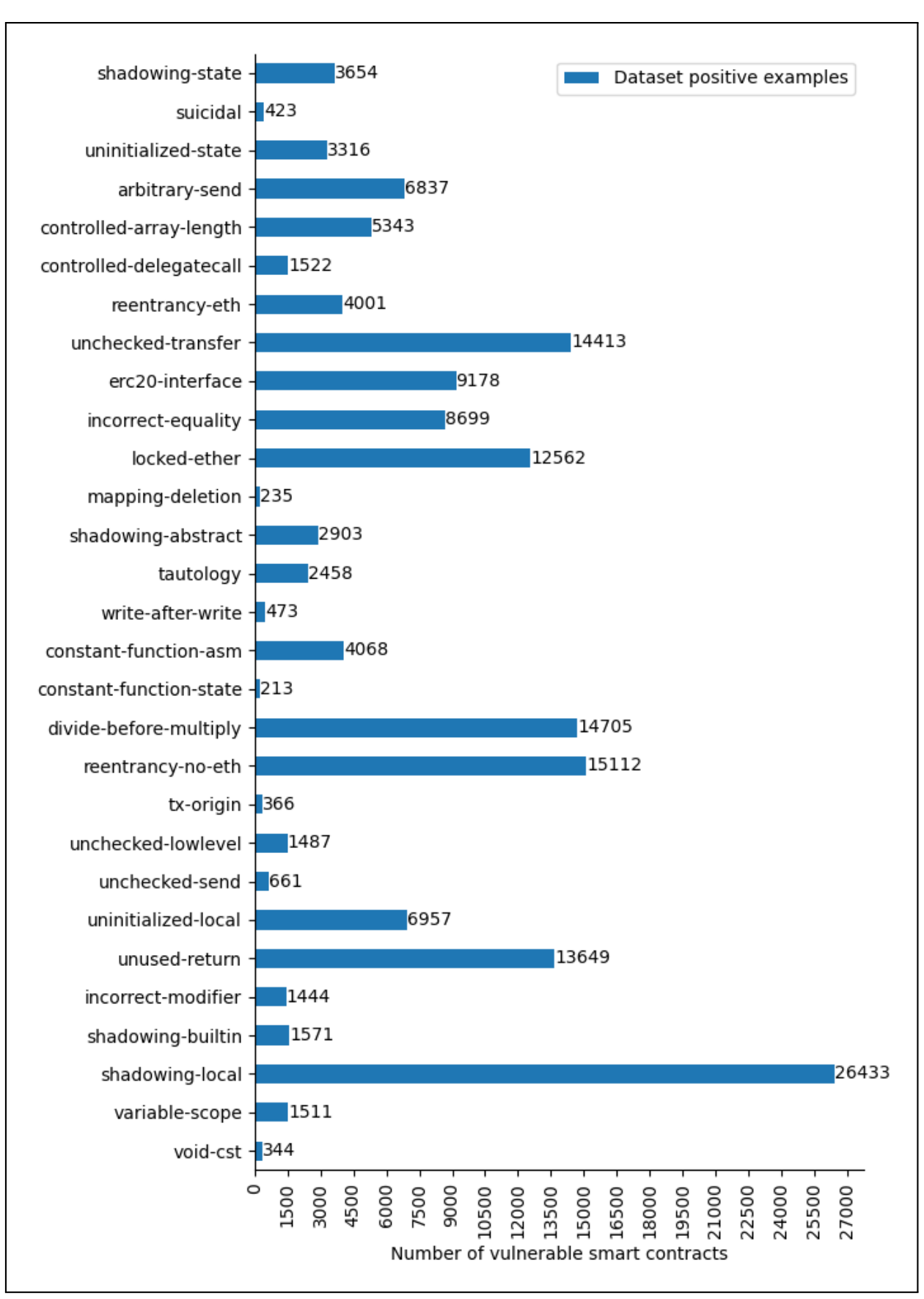

Figure Appx.12: Vulnerability Frequencies of EthereumSC Dataset

The benchmark is constructed as follows.

(a) We select five static analyzers: Oyente 0.2.7, Mythril 0.21.20, Osiris 0.0.1, Smartcheck 2.0, and Slither 0.8.0. The analyzers flag various vulnerabilities, but four are largely in common[6]: Re-entrancy, Timestamp-Dependency, Overflow-Underflow, and tx.origin.

(b) Out of our test set of 22,634 contracts in $EthereumSC_{large}$, we isolate the 553 contracts that are all considered safe for all four vulnerabilities, by all static analyzers that can detect them. We consider these 553 contracts to be the "negative ground truth."

(c) For each of these four vulnerabilities, SolidiFI has 40-45 code snippets exhibiting said vulnerability. By default, SolidiFI injects many vulnerabilities into target contracts. This makes the vulnerability detector's task too easy. To make the task more challenging, we modified SolidiFI to inject only a single randomly-chosen vulnerability into a given contract. Starting from a negative contract, we thus reach a "positive ground truth."

(d) For each vulnerability, we inject into each safe contract in a single randomly-chosen place of source code, yield-

ing $553 \times 4 = 2,212$ vulnerable contracts with unique bytecode for each contract after compilation. By the nature of the SolidiFI injection, none of these contracts have been seen before by any of the tools. Moreover, for a given vulnerability X, there will be 553 X-vulnerable contracts and 1,659 contracts that do not have X.

(e) We divide the contracts into three buckets: 60% for training, 20% for validation, and 20% for testing.

(f) This yields 111 contracts in the test set for each vulnerability (444 in total), with another 444 contracts in the validation set, and 1,324 in the training set.

# H The Zeus/eThor datasets (extending §4.1)

We also considered using an additional benchmark, first proposed to evaluate Zeus [13] and subsequently modified and used to evaluate eThor [21]. We found it challenging to use as a fair benchmark for the reasons we give below.

**Background of Zeus and eThor** Zeus [13] is a static analyzer that targets 7 smart contract vulnerabilities. Zeus focuses on soundness. Zeus achieves this by using the combination of abstract interpretation and software model checking. Zeus's analysis begins by converting the Solidity code into an abstract intermediate language, followed by a subsequent translation into LLVM bitcode. Analysis leverages well-established symbolic model checking tools for LLVM bitcode, facilitating a thorough examination of the smart contract. Unfortunately, Zeus is not publicly available, so we could not benchmark it.

eThor is based on an abstraction of the EVM bytecode semantics. The basic idea of abstract interpretation is to verify whether a program meets certain specific properties according to the approximation of the program's semantics. eThor builds on the work by Grishchenko et al. [8] that defines the full semantics of EVM bytecode, the researchers propose EtherTrust [9]. EtherTrust abstracts EVM bytecode into a series of Horn clauses and utilizes clause resolution to verify contract reachability. In response to the limitations of existing tools like EtherTrust [9], Zeus [13], NeuCheck [14], and MadMax [7] to provide provable soundness as contract analyzers, eThor [21] is an analyzer that offers provable soundness.

eThor has an available implementation and is proven sound against the EVM bytecode semantics defined by Grishchenko et al. [8]. Additionally, eThor can verify user-defined properties described in the *HoRSt* language, which simplifies the process of defining Horn clause-based contract abstractions.

**The Zeus and eThor Datasets** Although Zeus did not release their tool, they did release their dataset [26], showing both their ground truth labels as well as the result of running Zeus on the associated contracts. The authors write that ground truth was established by manual investigation [13], although a request to the authors for further detail on the standard by which labels were assigned yielded no response from 26 May 2023 through the publication date of this extended appendix on 19 June 2023 [10].

eThor took this dataset as a starting point. Since eThor only focuses on reentrancy, they ignored the other 6 vulnerabilities. eThor's authors found some data quality issues with Zeus's dataset as explained in the extended version of their paper [20, §D.5], trimming it down from 1,524 contracts to 720 contracts. 100 of these were trivially non-reentrant (no opcode that could cause reentrancy). 2 were out of scope for eThor because they included the `DELEGATECALL` or `CALLCODE` opcodes, so they removed them. The removed a further 13 because they were unable to reconstruct the CFG. Accordingly, they were left with 605 contracts. Most (508) have available source code on Etherscan. Contract sizes are mixed, up to 6,000 opcodes.

eThor provided a formal definition "single-entrancy," which is strictly more conservative than SWC-107. In other words, there are contracts SWC-107 labels as fixed/safe that eThor's ground truth considers unsafe [19]. eThor took the cleaned/trimmed reentrancy dataset from Zeus and relabelled them according to their single-entrancy definition. In most cases this was done using manual inspection of source code, although for bytecode-only contracts they used a decompiler and/or a comparison with similar contracts that did have available source [19]. The eThor team published this dataset [4].

**Our *Zeus/eThor$_{benchmark}$*** In turn, we analyzed Zeus's and eThor's datasets. We found two contracts lacked ground truth labels from both Zeus's and eThor's datasets. A further 20 contracts have ground truth labels in one dataset but not the other (10 per dataset). We removed these $2+10+10 = 22$ contracts also, leaving 583. We dub this dataset the *Zeus/eThor$_{benchmark}$* and make it public [1]. To facilitate comparison it includes the ground truth labels from both Zeus and eThor.

These ground truth labels differ *substantially*. Zeus assigns 21 Ground Positives and 562 Ground Negatives. Labelling exactly the same contracts, eThor assigns 184 GP and 399 GN. Moreover, although generally speaking eThor's labels seem more conservative (*i.e.*, many more GP), they are not a perfect superset: the contract at 0x6ff323e36bfdb20502b23780695f4e77e36cde95 was labelled at unsafe by Zeus but safe by eThor. The wide divergence in ground truth, the lack of clarity in Zeus's labelling strategy, and the divergence between single-entrancy and SWC-107 makes *Zeus/eThor$_{benchmark}$* difficult to interpret and use for fair benchmarking, which is why we do not include it in Figure 1, §4.4, §B, or Tables 4, 5, or 6.

**Our experiment** As part of our investigation into *Zeus/eThor$_{benchmark}$* we conducted the experiment summarized in Table Appx.7. Since 97 contracts do not have available source, and since the 508 contracts that do have source use very old versions of Solidity (before 0.4.0, circa 2016), we restricted ourselves to the six analyzers that handle bytecode. As explained above, we are unsure of Zeus's labelling strategy, so we report only the results for eThor's ground truth.

We divide the results into two big columns, one for the contracts ground-labeled as positive (GP) and the second for those ground-labeled as negative (GN). Within both big columns, "Exp" tracks the number of contracts for which analysis failed (*e.g.*, timeouts). In the GP big column, "FN" gives the number of False Negatives, *i.e.* GPs incorrectly reported as safe; and "TP" gives the number of True Positives, *i.e.* GPs correctly reported as unsafe. In the GN big column, "FP" gives the number of False Positives, *i.e.* GNs incorrectly reported as unsafe; and "TN" gives the number of True Negatives, *i.e.* GNs correctly reported as safe.

Since Zeus is not publicly available, our data for that tool is determined only from the labels given in eThor's dataset [4], although we did some spot checking (21 contracts, 3.6% of the total) to ensure that these labels had been correctly copied from Zeus's dataset [26].

We report eThor twice: once from the dataset published by the authors ("eThor"), and a second time where we ran eThor ourselves ("eThor by us"). The numbers differ because we had many more contracts time out. Our test machine had 12 cores running at 3.2 GHz and 16 GB of RAM, whereas theirs had 24 cores running at 2.8 GHz and 150 GB of RAM. We used a 30 minute timeout, whereas they used only a 10 minute timeout, but it was not enough to overcome the raw hardware advantage. When our tests did terminate, we verified that the labels were the same as those reported by eThor [4].

The results from the other tools (Oyente, Osiris, Mythril, and DLVA) were from running these tools ourselves. In all cases we used the same tool versions reported in Table 3 that were used in our main experiments.

The results are generally in line with our expectations, given that single-entrancy is strictly more conservative than SWC-107. All tools except eThor report huge numbers of false negatives (*i.e.*, significant unsoundness with respect to single-entrancy). This is basically the converse of eThor's performance on our main experiments, which concluded that eThor's average False Positive Rate was 79.8% vs. the average competitor's 12.1% and DLVA's 0.6% (Figure 1).

Looking slightly deeper into the data, there is substantial overlap between the labels of Oyente, Osiris, and DLVA. Consider the 184 GPs. There are 155 contracts (84.2%) that all three label as safe, but which eThor's single-entrancy ground truth considers unsafe (*i.e.*, false negatives)[4]. For the remaining 29 GP contracts, there is disturbingly little overlap between the tools: in fact, no contract is considered unsafe by all six tools. Without (at the very least) manually relabelling this benchmark with respect to SWC-107, it is very difficult to say if the reasons for these results are due to mistakes made by these tools, differences in their underlying definitions for reentrancy (beyond single-entrancy's strict conservatism as compared to SWC-107), or other factors.

[4] Of these 155, Zeus considers 142 safe. Mythril's results are harder to interpret due to the large number of timeouts, but 51 of its 55 reported false negatives are in this subset of 155 contracts.

Table Appx.7: Comparison of DLVA vs. state-of-the-art tools based on eThor's ground-truth labels.

| **Analyzer** | **eThor's labels** [1] | | | | | |
|---|---|---|---|---|---|---|
| | 184 GP | | | 399 GN | | |
| | Exp | FN | TP | Exp | FP | TN |
| **Oyente** [15] | 5 | 174 | 5 | 0 | 0 | 399 |
| **Osiris** [24] | 0 | 170 | 14 | 2 | 1 | 396 |
| **Mythril** [17] | 41 | 55 | 88 | 85 | 3 | 311 |
| **Zeus** [13] | 0 | 164 | 20 | 0 | 1 | 398 |
| **eThor** [21] | 1 | 0 | 183 | 14 | 96 | 289 |
| **eThor by us** | 26 | 0 | 158 | 55 | 82 | 262 |
| **DLVA** | 1 | 174 | 9 | 2 | 11 | 386 |

# I Finding The Right Machine Learning Model for Node Feature Extraction (extending §4.2)

Dealing with EVM bytecode instructions as text data is problematic, since machine learning models can not interpret text directly, and require numerical feature vectors. Our dataset contains approximately 714 million instructions. We tried to apply word embedding methods such as *fastText* [2] and *word2vec* [16] to learn the vector representations of EVM instructions. *fastText* outperforms *word2vec* and reduces the loss to 2.3% compared to 30% in *word2vec*. Both require an appropriate composition function to aggregate each basic block instructions (CFG node) into a single vector.

To aggregate all instructions of each CFG node $n_i$ into a single vector $\vec{x_i}$, we tried to apply Recurrent Neural Networks (RNNs) as a composition function which distill variable-length sequences of vector-represented data into a single vector that summarizes the sequence. RNNs are powerful machine learning models adapted to sequence data and operate over sequences of vectors one by one. RNNs can use their internal state (memory) to process variable length sequences of instructions. RNNs are great when we are dealing with short-term dependencies, but fail to understand and remember the context behind long-term sequences. This is because RNNs suffer from the vanishing gradient problem that happens when gradient (values used to update a neural networks weights) shrinks as it back propagates through time of processing long sequences. RNNs layer that gets an extremely small gradient update doesn't contribute too much learning. So because some layers stop learning, RNNs can forget what it seen in longer sequences. Long Short Term Memory Networks (LSTMs) [11, 5] have an edge over RNNs because of their property of selectively remembering patterns for long duration of time and the ability of learning long-term dependencies. LSTMs have the ability to remove or add information to the cell state using gates. LSTMs gates can learn which data in a sequence is important to keep or throw away to keep the important features in long sequences.

We tried to apply 2-layer of Bidirectional Long Short-Term Memory (BiLSTMs) [22], as shown in Figure Appx.13. BiLSTM consists of two LSTMs, one taking the input in a forward direction, and the other in a backwards direction. BiLSTMs can learn and achieve outstanding performance and accuracy on many hard sequential learning problems. The BiLSTMs model is trained end-to-end, takes in input all the instruction embedding vectors in order, i.e. $(\overrightarrow{i_1}, ..., \overrightarrow{i_m})$ and generates m outputs and m hidden states $(\overrightarrow{h_{(1)}}, ..., \overrightarrow{h_{(m)}})$. The final feature vector $\overrightarrow{x_i}$ for each node is simply the last hidden state $\overrightarrow{h_{(m)}}$.

In our proposed architecture, DLVA, we utilized the Universal Sentence Encoder (USE), which has demonstrated superior performance compared to *fastText* followed by BiLSTMs. We first use *fastText* to learn the representation of EVM instructions, and then employ BiLSTMs to aggregate the instructions of each node, taking approximately 6.22 seconds per contract. Alternatively, by utilizing USE, we can transform these instructions in just 0.3 seconds per contract.

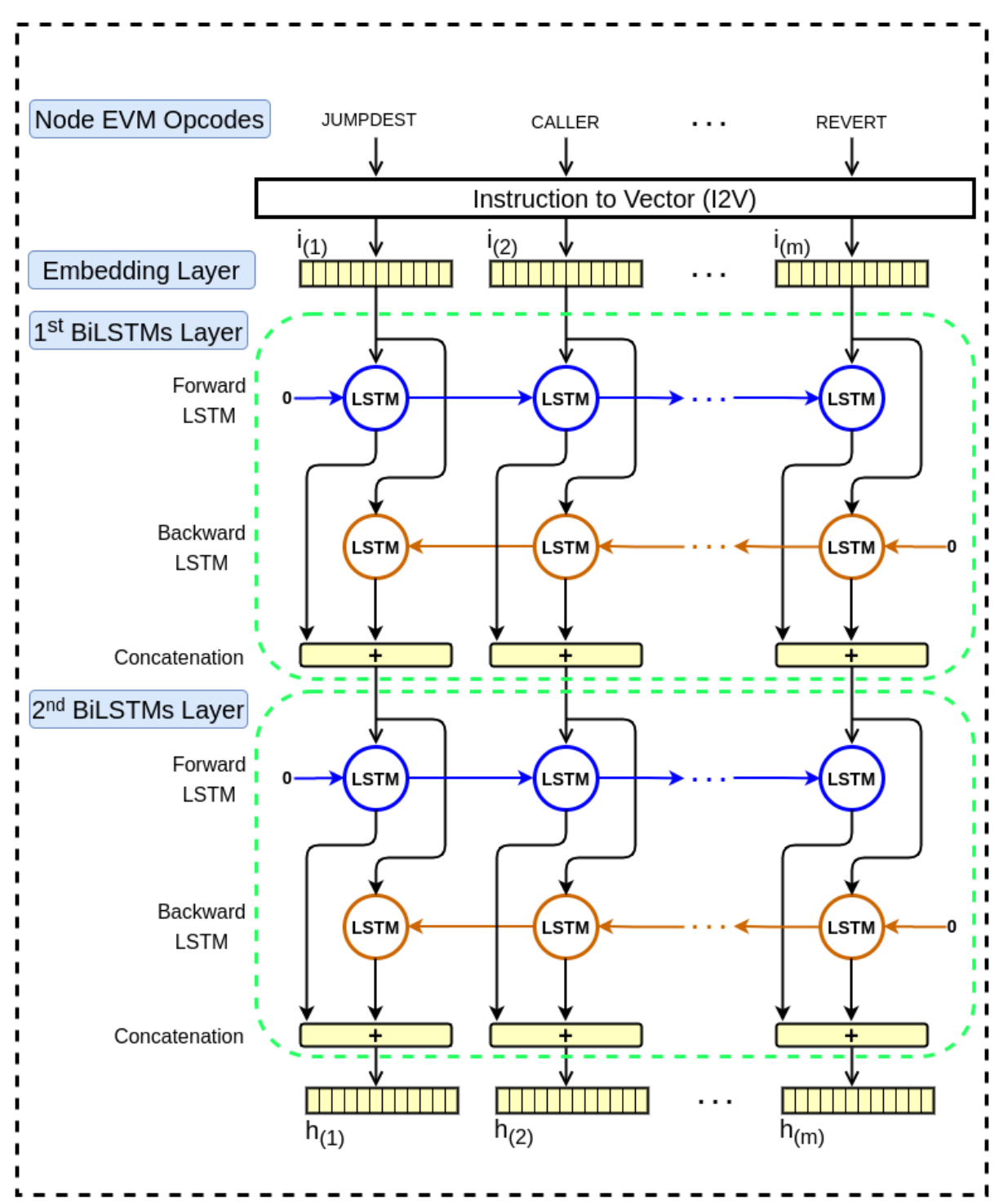

Figure Appx.13: Node Instructions Composition using Bidirectional Long Short-Term Memory.

## J Comparing SC2V to Alternatives (extending §4.2)

Figure Appx.14 contains the breakdown of Figure 5 by individual vulnerabilities. SC2V wins outright for Reentrancy and Overflow/Underflow; ties for tx.origin; and comes in second for Timestamp-Dependency.

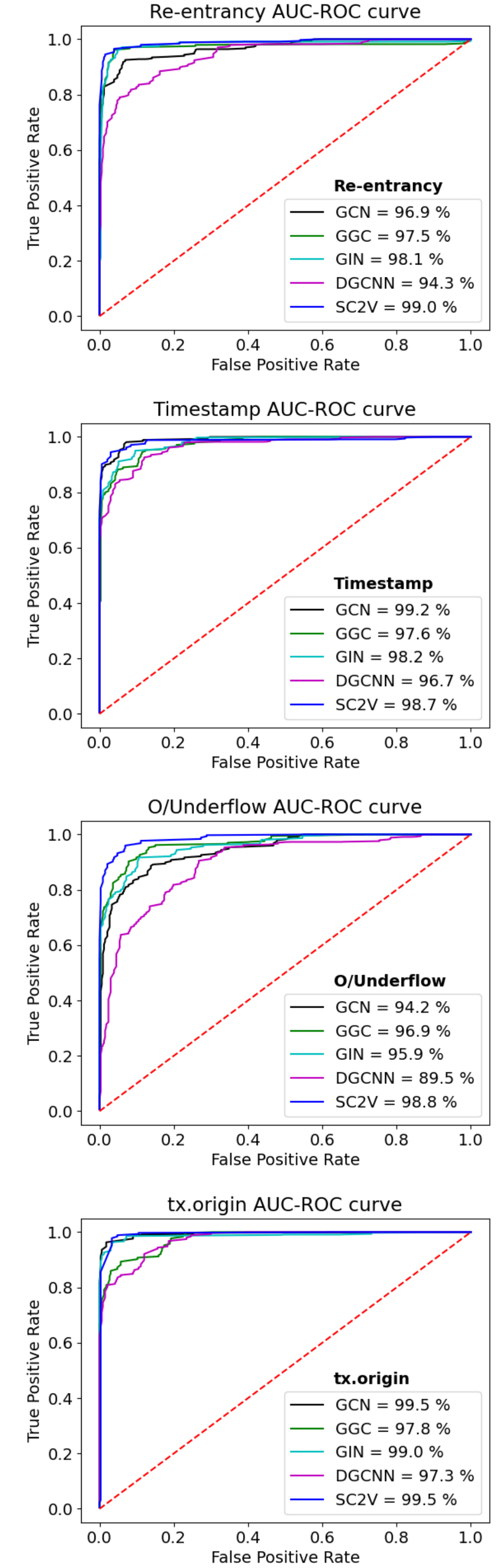

Figure Appx.14: Evaluating SC2V vs. state-of-the-art GNNs

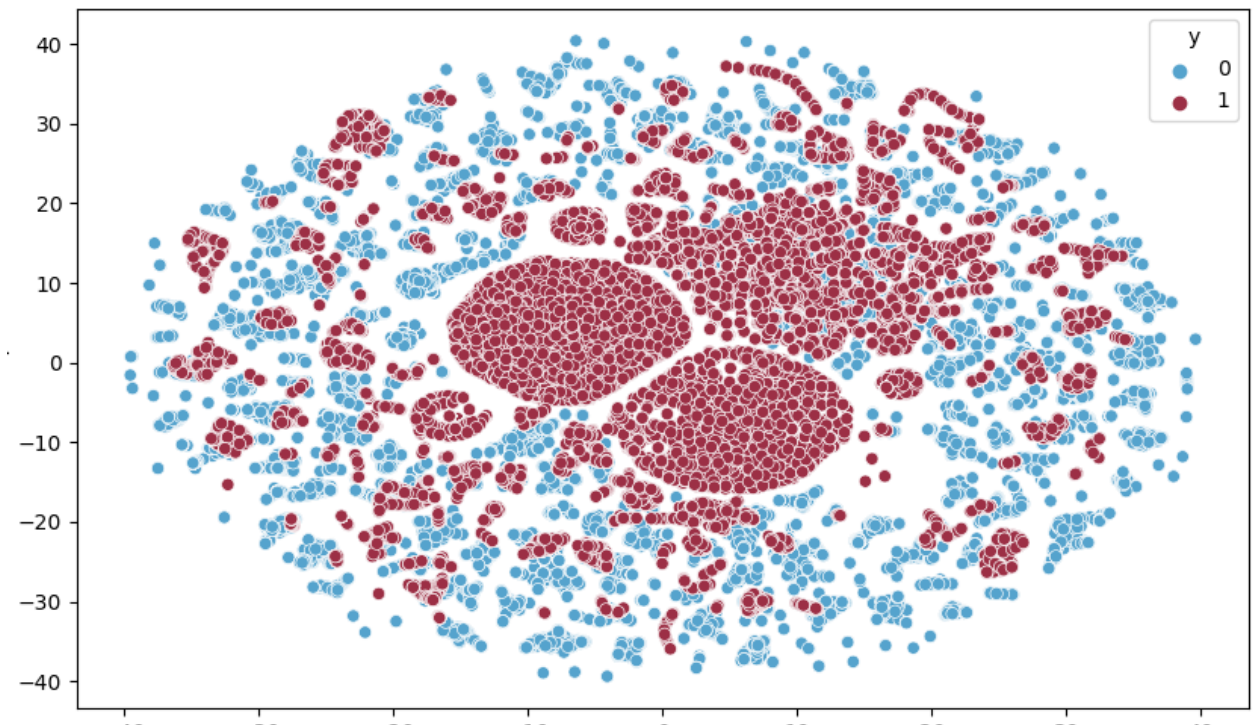

Figure Appx.15: t-SNE-Embeddings for the “unchecked-lowlevel” Vulnerability.

The point of SC2V is to Euclidian-cluster smart contracts that share a specific vulnerability distinctly from contracts that are safe from that vulnerability. The underlying vectors for large contracts have 4,128 dimensions. We use SMOTEENN (combining over- and under-sampling using SMOTE and Edited Nearest Neighbours) to balance positive and negative examples. Afterward, we employ t-Distributed Stochastic Neighbor Embedding (t-SNE) to compress into two dimensions for visualization purposes. Figure Appx.15 does just this for the “unchecked-lowlevel” vulnerability, with Slither-labeled vulnerable contracts in red and safe contracts in blue. Even given this compression, the clustering effect is apparent.

## K Comparing the Core Classifier vs Machine Learning Classifiers (extending §4.2)

It is not easy to find a machine learning model that performs better than DLVA’s Core Classifier. To demonstrate this, we benchmark the CC against ten commonly used machine learning supervised binary classifiers[5]: Logistic Regression (LR), Random Forest (RF), Stochastic gradient descent (SGD), Support Vector Machines (SVM), Extra Trees (ET), Multi-layer Perceptron (MLP), K-Nearest Neighbour (KNN), Gradient Boosting (GB), XGBoost (XGB), and AdaBoost (AB). To get a fair fight, all ten ML classifiers were well-tuned: each was independently trained with three-fold cross validation for all 29 vulnerabilities. That is, in total we trained and evaluated 290 competitors. In addition, we built a “meta-competitor” that used majority voting between the competitors to further increase accuracy. The individual models have “weak vulnerabilities,” for which they do not shine; voting smooths weaknesses out.

Figure 6 presents the results of our competition. We measure three key statistics: on the left, accuracy (higher is better); in the middle, the True Positive Rate (higher is better); and at the right, the False Positive Rate (lower is better). DLVA’s CC is more accurate than every other model, for every test. Moreover, the CC usually enjoys the highest/best TPR (or close), and the lowest/best FPR (or close). On a few tests the CC’s TPR is uninspiring: uninitialized-state and write-after-write are the most challenging. Fortunately, on those difficult vulnerabilities, the CC’s excellent FPR comes to the rescue. Conversely, the CC’s FPR is uninspiring for constant-function-state; happily, it leads the pack on the associated TPR.

The voting meta-competitor also performs well: well above average accuracy, for every test. But DLVA’s CC is better: it reduces the size of the error set—the set of contracts for which a given classifier goofs—by an average of 36%.

We put Table Appx.8, which examines the competition from a “post-hoc” viewpoint. That is, while the meta-competitor does not know which of the 10 machine learning models will be the most accurate for a given vulnerability, in Table Appx.8 we get to bet on the winners after the race has been run! Despite this advantage, DLVA’s CC reduces the average size of the error set by 30%.

DLVA-CC achieved an average accuracy of 80.0%, surpassing the average competitors accuracy of 68.94%. This indicates that DLVA’s CC effectively reduces the average size of the error set by approximately $36\% \approx 1 - \frac{100\text{ - }80.0}{100\text{ - }68.94} * 100$. In addition, DLVA-CC even outperforms the highest competitor’s accuracy of 71.6%. This indicates that DLVA’s CC effectively reduces the typical size of the error set by approximately $30\% \approx 1 - \frac{100\text{ - }80.0}{100\text{ - }71.6} * 100$.

## L Detailed Tables for Our Experiments in §4.3

Table Appx.9 showcases the performance of the CC when runing on the entire test set. Table Appx.10 presents the results obtained from the Sibling Detector (SD), shedding light on its performance, on the “easy” portion of the test set. Table Appx.11 highlights the performance of the DLVA Core Classifier (CC) on the “hard” portion of the test set. Finally, Table Appx.12 combines both components of DLVA (SD and CC). All of these tables cover the version of DLVA that was trained on $EthereumSC_{large}$.

Table Appx.13 presents the performance results of DLVA trained on $EthereumSC_{small}$. Recall from Table 1 that we use only 21 vulnerabilities due to the relatively small number of positive occurrences in contracts in this dataset. For the same reason, we do not use the Sibling Detector.

## Appendix References

[5] Machine Learning Classifiers in Python: https://scikit-learn.org/stable/supervised_learning.html#supervised-learning

Table Appx.8: Best of the ten commonly used machine learning supervised binary classifiers results

| **Vulnerability** | **Classifier** | **Test size** | **TP** | **FP** | **TN** | **FN** | **Accuracy** | **TPR** | **TNR** | **FPR** | **FNR** | **AUC** |
|---|---|---|---|---|---|---|---|---|---|---|---|---|
| **shadowing-state** | **MLP** | **10037** | 290 | 1525 | 8038 | 184 | 72.6% | 61.2% | 84.1% | 15.9% | 38.8% | 78.6% |
| **suicidal** | **SVM** | **10037** | 58 | 3487 | 6482 | 10 | 75.2% | 85.3% | 65.0% | 35.0% | 14.7% | 78.3% |
| **uninitialized-state** | **MLP** | **10037** | 198 | 3851 | 5854 | 134 | 60.0% | 59.6% | 60.3% | 39.7% | 40.4% | 62.9% |
| **arbitrary-send** | **MLP** | **10037** | 653 | 2394 | 6755 | 235 | 73.7% | 73.5% | 73.8% | 26.2% | 26.5% | 80.4% |
| **controlled-array-length** | **XGB** | **10037** | 526 | 2688 | 6653 | 170 | 73.4% | 75.6% | 71.2% | 28.8% | 24.4% | 80.1% |
| **controlled-delegatecall** | **SVM** | **10037** | 54 | 3059 | 6909 | 15 | 73.8% | 78.3% | 69.3% | 30.7% | 21.7% | 76.1% |
| **reentrancy-eth** | **MLP** | **10037** | 377 | 2701 | 6842 | 117 | 74.0% | 76.3% | 71.7% | 28.3% | 23.7% | 81.1% |
| **reentrancy-no-eth** | **ET** | **10037** | 1373 | 2438 | 5812 | 414 | 73.6% | 76.8% | 70.4% | 29.6% | 23.2% | 79.6% |
| **unchecked-transfer** | **SVM** | **10037** | 1428 | 2672 | 5647 | 290 | 75.5% | 83.1% | 67.9% | 32.1% | 16.9% | 81.2% |
| **erc20-interface** | **MLP** | **10037** | 522 | 3138 | 6157 | 220 | 68.3% | 70.4% | 66.2% | 33.8% | 29.6% | 73.5% |
| **incorrect-equality** | **SVM** | **10037** | 932 | 3161 | 5614 | 330 | 68.9% | 73.9% | 64.0% | 36.0% | 26.1% | 73.6% |
| **locked-ether** | **ET** | **10037** | 513 | 3881 | 5445 | 198 | 65.3% | 72.2% | 58.4% | 41.6% | 27.8% | 69.3% |
| **mapping-deletion** | **ET** | **10037** | 22 | 1954 | 8052 | 9 | 75.7% | 71.0% | 80.5% | 19.5% | 29.0% | 79.5% |
| **shadowing-abstract** | **XGB** | **10037** | 192 | 1997 | 7776 | 72 | 76.1% | 72.7% | 79.6% | 20.4% | 27.3% | 82.6% |
| **tautology** | **MLP** | **10037** | 223 | 3098 | 6620 | 96 | 69.0% | 69.9% | 68.1% | 31.9% | 30.1% | 74.6% |
| **write-after-write** | **KNN** | **10037** | 62 | 4004 | 5949 | 22 | 66.8% | 73.8% | 59.8% | 40.2% | 26.2% | 67.5% |
| **constant-function-asm** | **MLP** | **10037** | 423 | 2826 | 6723 | 65 | 78.5% | 86.7% | 70.4% | 29.6% | 13.3% | 84.5% |
| **constant-function-state** | **AB** | **10037** | 24 | 1906 | 8100 | 7 | 79.2% | 77.4% | 81.0% | 19.0% | 22.6% | 84.2% |
| **divide-before-multiply** | **SVM** | **10037** | 1361 | 2909 | 5268 | 499 | 68.8% | 73.2% | 64.4% | 35.6% | 26.8% | 73.0% |
| **tx-origin** | **MLP** | **10037** | 37 | 3487 | 6506 | 7 | 74.6% | 84.1% | 65.1% | 34.9% | 15.9% | 77.3% |
| **unchecked-lowlevel** | **LR** | **10037** | 53 | 2498 | 7469 | 17 | 75.3% | 75.7% | 74.9% | 25.1% | 24.3% | 80.3% |
| **unchecked-send** | **XGB** | **10037** | 69 | 3602 | 6356 | 10 | 75.6% | 87.3% | 63.8% | 36.2% | 12.7% | 79.5% |
| **uninitialized-local** | **SVM** | **10037** | 734 | 4161 | 4894 | 248 | 64.4% | 74.7% | 54.0% | 46.0% | 25.3% | 68.1% |
| **unused-return** | **XGB** | **10037** | 1378 | 2884 | 5494 | 281 | 74.3% | 83.1% | 65.6% | 34.4% | 16.9% | 80.1% |
| **incorrect-modifier** | **MLP** | **10037** | 104 | 3353 | 6533 | 47 | 67.5% | 68.9% | 66.1% | 33.9% | 31.1% | 71.9% |
| **shadowing-builtin** | **MLP** | **10037** | 115 | 2706 | 7172 | 44 | 72.5% | 72.3% | 72.6% | 27.4% | 27.7% | 77.6% |
| **shadowing-local** | **MLP** | **10037** | 1534 | 3025 | 4624 | 854 | 62.3% | 64.2% | 60.5% | 39.5% | 35.8% | 68.1% |
| **variable-scope** | **MLP** | **10037** | 194 | 3413 | 6382 | 48 | 72.7% | 80.2% | 65.2% | 34.8% | 19.8% | 79.0% |
| **void-cst** | **AB** | **10037** | 35 | 3747 | 6243 | 12 | 68.5% | 74.5% | 62.5% | 37.5% | 25.5% | 73.2% |

Table Appx.9: Task CC-only: use DLVA's core classifier by itself for the entire test set in $EthereumSC_{large}$

| **Vulnerability** | **Test size** | **TP** | **FP** | **TN** | **FN** | **Accuracy** | **TPR** | **TNR** | **FPR** | **FNR** |
|---|---|---|---|---|---|---|---|---|---|---|
| **shadowing-state** | **22634** | 537 | 2715 | 19189 | 193 | 80.6% | 73.6% | 87.6% | 12.4% | 26.4% |
| **suicidal** | **22634** | 77 | 1456 | 21094 | 7 | 92.6% | 91.7% | 93.5% | 6.5% | 8.3% |
| **uninitialized-state** | **22634** | 439 | 4599 | 17398 | 198 | 74.0% | 68.9% | 79.1% | 20.9% | 31.1% |
| **arbitrary-send** | **22634** | 1227 | 3715 | 17573 | 119 | 86.9% | 91.2% | 82.6% | 17.4% | 8.8% |
| **controlled-array-length** | **22634** | 939 | 2481 | 19117 | 97 | 89.6% | 90.6% | 88.5% | 11.5% | 9.4% |
| **controlled-delegatecall** | **22634** | 266 | 1148 | 21191 | 29 | 92.5% | 90.2% | 94.9% | 5.1% | 9.8% |
| **reentrancy-eth** | **22634** | 702 | 3816 | 18017 | 99 | 85.1% | 87.6% | 82.5% | 17.5% | 12.4% |
| **reentrancy-no-eth** | **22634** | 2509 | 3153 | 16528 | 444 | 84.5% | 85.0% | 84.0% | 16.0% | 15.0% |
| **unchecked-transfer** | **22634** | 2577 | 1948 | 17856 | 253 | 90.6% | 91.1% | 90.2% | 9.8% | 8.9% |
| **erc20-interface** | **22634** | 1598 | 1845 | 19043 | 148 | 91.3% | 91.5% | 91.2% | 8.8% | 8.5% |
| **incorrect-equality** | **22634** | 1561 | 5489 | 15374 | 210 | 80.9% | 88.1% | 73.7% | 26.3% | 11.9% |
| **locked-ether** | **22634** | 2092 | 2366 | 17718 | 458 | 85.1% | 82.0% | 88.2% | 11.8% | 18.0% |
| **mapping-deletion** | **22634** | 33 | 1869 | 20724 | 8 | 86.1% | 80.5% | 91.7% | 8.3% | 19.5% |
| **shadowing-abstract** | **22634** | 570 | 1126 | 20885 | 53 | 93.2% | 91.5% | 94.9% | 5.1% | 8.5% |
| **tautology** | **22634** | 391 | 4701 | 17451 | 91 | 80.0% | 81.1% | 78.8% | 21.2% | 18.9% |
| **write-after-write** | **22634** | 73 | 5309 | 17228 | 24 | 75.9% | 75.3% | 76.4% | 23.6% | 24.7% |
| **constant-function-asm** | **22634** | 794 | 1889 | 19928 | 23 | 94.3% | 97.2% | 91.3% | 8.7% | 2.8% |
| **constant-function-state** | **22634** | 33 | 2864 | 19726 | 11 | 81.2% | 75.0% | 87.3% | 12.7% | 25.0% |
| **divide-before-multiply** | **22634** | 2468 | 2935 | 16793 | 438 | 85.0% | 84.9% | 85.1% | 14.9% | 15.1% |
| **tx-origin** | **22634** | 55 | 4191 | 18381 | 7 | 85.1% | 88.7% | 81.4% | 18.6% | 11.3% |
| **unchecked-lowlevel** | **22634** | 276 | 611 | 21732 | 15 | 96.1% | 94.8% | 97.3% | 2.7% | 5.2% |
| **unchecked-send** | **22634** | 116 | 3718 | 18792 | 8 | 88.5% | 93.5% | 83.5% | 16.5% | 6.5% |
| **uninitialized-local** | **22634** | 1162 | 5170 | 16124 | 178 | 81.2% | 86.7% | 75.7% | 24.3% | 13.3% |
| **unused-return** | **22634** | 1947 | 3991 | 16354 | 342 | 82.7% | 85.1% | 80.4% | 19.6% | 14.9% |
| **incorrect-modifier** | **22634** | 219 | 2875 | 19509 | 31 | 87.4% | 87.6% | 87.2% | 12.8% | 12.4% |
| **shadowing-builtin** | **22634** | 274 | 1174 | 21145 | 41 | 90.9% | 87.0% | 94.7% | 5.3% | 13.0% |
| **shadowing-local** | **22634** | 4595 | 3416 | 13987 | 636 | 84.1% | 87.8% | 80.4% | 19.6% | 12.2% |
| **variable-scope** | **22634** | 250 | 3457 | 18874 | 53 | 83.5% | 82.5% | 84.5% | 15.5% | 17.5% |
| **void-cst** | **22634** | 59 | 3589 | 18977 | 9 | 85.4% | 86.8% | 84.1% | 15.9% | 13.2% |

Table Appx.10: Task SD-easy: use the Sibling Detector for the “easy” contracts in $EthereumSC_{large}$

| **Vulnerability** | **Test size** ($d <= 0.1$) | **TP** | **FP** | **TN** | **FN** | **Accuracy** | **TPR** | **TNR** | **FPR** | **FNR** | **Test size** ($d > 0.1$) |
|---|---|---|---|---|---|---|---|---|---|---|---|
| **shadowing-state** | **12597** | 244 | 6 | 12335 | 12 | 97.6% | 95.3% | 100.0% | 0.0% | 4.7% | **10037** |
| **suicidal** | **12597** | 14 | 0 | 12581 | 2 | 93.8% | 87.5% | 100.0% | 0.0% | 12.5% | **10037** |
| **uninitialized-state** | **12597** | 273 | 6 | 12286 | 32 | 94.7% | 89.5% | 100.0% | 0.0% | 10.5% | **10037** |
| **arbitrary-send** | **12597** | 444 | 12 | 12127 | 14 | 98.4% | 96.9% | 99.9% | 0.1% | 3.1% | **10037** |
| **controlled-array-length** | **12597** | 316 | 22 | 12236 | 23 | 96.5% | 93.2% | 99.8% | 0.2% | 6.8% | **10037** |
| **controlled-delegatecall** | **12597** | 219 | 2 | 12369 | 7 | 98.4% | 96.9% | 100.0% | 0.0% | 3.1% | **10037** |
| **reentrancy-eth** | **12597** | 290 | 11 | 12279 | 17 | 97.2% | 94.5% | 99.9% | 0.1% | 5.5% | **10037** |
| **reentrancy-no-eth** | **12597** | 1126 | 35 | 11396 | 40 | 98.1% | 96.6% | 99.7% | 0.3% | 3.4% | **10037** |
| **unchecked-transfer** | **12597** | 1081 | 39 | 11446 | 31 | 98.4% | 97.2% | 99.7% | 0.3% | 2.8% | **10037** |
| **erc20-interface** | **12597** | 1000 | 9 | 11584 | 4 | 99.8% | 99.6% | 99.9% | 0.1% | 0.4% | **10037** |
| **incorrect-equality** | **12597** | 483 | 37 | 12052 | 25 | 97.4% | 95.1% | 99.7% | 0.3% | 4.9% | **10037** |
| **locked-ether** | **12597** | 1810 | 24 | 10735 | 28 | 99.1% | 98.5% | 99.8% | 0.2% | 1.5% | **10037** |
| **mapping-deletion** | **12597** | 10 | 0 | 12587 | 0 | 100.0% | 100.0% | 100.0% | 0.0% | 0.0% | **10037** |
| **shadowing-abstract** | **12597** | 357 | 5 | 12233 | 2 | 99.7% | 99.4% | 100.0% | 0.0% | 0.6% | **10037** |
| **tautology** | **12597** | 158 | 3 | 12432 | 4 | 98.8% | 97.5% | 100.0% | 0.0% | 2.5% | **10037** |
| **write-after-write** | **12597** | 11 | 3 | 12581 | 2 | 92.3% | 84.6% | 100.0% | 0.0% | 15.4% | **10037** |
| **constant-function-asm** | **12597** | 320 | 6 | 12263 | 8 | 98.8% | 97.6% | 100.0% | 0.0% | 2.4% | **10037** |
| **constant-function-state** | **12597** | 12 | 0 | 12584 | 1 | 96.2% | 92.3% | 100.0% | 0.0% | 7.7% | **10037** |
| **divide-before-multiply** | **12597** | 1008 | 26 | 11526 | 37 | 98.1% | 96.5% | 99.8% | 0.2% | 3.5% | **10037** |
| **tx-origin** | **12597** | 18 | 0 | 12579 | 0 | 100.0% | 100.0% | 100.0% | 0.0% | 0.0% | **10037** |
| **unchecked-lowlevel** | **12597** | 220 | 0 | 12376 | 1 | 99.8% | 99.5% | 100.0% | 0.0% | 0.5% | **10037** |
| **unchecked-send** | **12597** | 42 | 0 | 12553 | 2 | 97.7% | 95.5% | 100.0% | 0.0% | 4.5% | **10037** |
| **uninitialized-local** | **12597** | 343 | 9 | 12230 | 15 | 97.9% | 95.8% | 99.9% | 0.1% | 4.2% | **10037** |
| **unused-return** | **12597** | 608 | 24 | 11943 | 22 | 98.2% | 96.5% | 99.8% | 0.2% | 3.5% | **10037** |
| **incorrect-modifier** | **12597** | 88 | 9 | 12489 | 11 | 94.4% | 88.9% | 99.9% | 0.1% | 11.1% | **10037** |
| **shadowing-builtin** | **12597** | 151 | 5 | 12436 | 5 | 98.4% | 96.8% | 100.0% | 0.0% | 3.2% | **10037** |
| **shadowing-local** | **12597** | 2780 | 149 | 9607 | 61 | 98.2% | 97.9% | 98.5% | 1.5% | 2.1% | **10037** |
| **variable-scope** | **12597** | 55 | 6 | 12531 | 5 | 95.8% | 91.7% | 100.0% | 0.0% | 8.3% | **10037** |
| **void-cst** | **12597** | 17 | 2 | 12574 | 4 | 90.5% | 81.0% | 100.0% | 0.0% | 19.0% | **10037** |

Table Appx.11: Task CC-hard: use the Core Classifier on the “hard” contracts in *EthereumSC$_{large}$*

| Vulnerability | Test Size | TP | FP | TN | FN | Accuracy | TPR | TNR | FPR | FNR | AUC |
|---|---|---|---|---|---|---|---|---|---|---|---|
| **shadowing-state** | **10037** | 334 | 1660 | 7903 | 140 | 76.6% | 70.5% | 82.6% | 17.4% | 29.5% | 83.0% |
| **suicidal** | **10037** | 62 | 1063 | 8906 | 6 | 90.3% | 91.2% | 89.3% | 10.7% | 8.8% | 94.6% |
| **uninitialized-state** | **10037** | 190 | 2586 | 7119 | 142 | 65.3% | 57.2% | 73.4% | 26.6% | 42.8% | 68.4% |
| **arbitrary-send** | **10037** | 759 | 2449 | 6700 | 129 | 79.4% | 85.5% | 73.2% | 26.8% | 14.5% | 86.5% |
| **controlled-array-length** | **10037** | 589 | 1378 | 7963 | 107 | 84.9% | 84.6% | 85.2% | 14.8% | 15.4% | 91.8% |
| **controlled-delegatecall** | **10037** | 57 | 2085 | 7883 | 12 | 80.8% | 82.6% | 79.1% | 20.9% | 17.4% | 88.4% |
| **reentrancy-eth** | **10037** | 413 | 2366 | 7177 | 81 | 79.4% | 83.6% | 75.2% | 24.8% | 16.4% | 86.1% |
| **reentrancy-no-eth** | **10037** | 1430 | 2071 | 6179 | 357 | 77.5% | 80.0% | 74.9% | 25.1% | 20.0% | 85.7% |
| **unchecked-transfer** | **10037** | 1516 | 1361 | 6958 | 202 | 85.9% | 88.2% | 83.6% | 16.4% | 11.8% | 91.8% |
| **erc20-interface** | **10037** | 613 | 1287 | 8008 | 129 | 84.4% | 82.6% | 86.2% | 13.8% | 17.4% | 92.3% |
| **incorrect-equality** | **10037** | 990 | 2685 | 6090 | 272 | 73.9% | 78.4% | 69.4% | 30.6% | 21.6% | 81.5% |
| **locked-ether** | **10037** | 541 | 2572 | 6754 | 170 | 74.3% | 76.1% | 72.4% | 27.6% | 23.9% | 81.4% |
| **mapping-deletion** | **10037** | 23 | 1336 | 8670 | 8 | 80.4% | 74.2% | 86.6% | 13.4% | 25.8% | 82.8% |
| **shadowing-abstract** | **10037** | 224 | 917 | 8856 | 40 | 87.7% | 84.8% | 90.6% | 9.4% | 15.2% | 94.1% |
| **tautology** | **10037** | 238 | 2925 | 6793 | 81 | 72.3% | 74.6% | 69.9% | 30.1% | 25.4% | 80.0% |
| **write-after-write** | **10037** | 61 | 3388 | 6565 | 23 | 69.3% | 72.6% | 66.0% | 34.0% | 27.4% | 75.8% |
| **constant-function-asm** | **10037** | 448 | 902 | 8647 | 40 | 91.2% | 91.8% | 90.6% | 9.4% | 8.2% | 97.1% |
| **constant-function-state** | **10037** | 27 | 2711 | 7295 | 4 | 80.0% | 87.1% | 72.9% | 27.1% | 12.9% | 87.2% |
| **divide-before-multiply** | **10037** | 1496 | 1800 | 6377 | 364 | 79.2% | 80.4% | 78.0% | 22.0% | 19.6% | 87.0% |
| **tx-origin** | **10037** | 37 | 2606 | 7387 | 7 | 79.0% | 84.1% | 73.9% | 26.1% | 15.9% | 83.7% |
| **unchecked-lowlevel** | **10037** | 65 | 925 | 9042 | 5 | 91.8% | 92.9% | 90.7% | 9.3% | 7.1% | 96.3% |
| **unchecked-send** | **10037** | 72 | 2387 | 7571 | 7 | 83.6% | 91.1% | 76.0% | 24.0% | 8.9% | 89.2% |
| **uninitialized-local** | **10037** | 736 | 2700 | 6355 | 246 | 72.6% | 74.9% | 70.2% | 29.8% | 25.1% | 79.6% |
| **unused-return** | **10037** | 1332 | 2205 | 6173 | 327 | 77.0% | 80.3% | 73.7% | 26.3% | 19.7% | 84.4% |
| **incorrect-modifier** | **10037** | 125 | 1912 | 7974 | 26 | 81.7% | 82.8% | 80.7% | 19.3% | 17.2% | 88.7% |
| **shadowing-builtin** | **10037** | 135 | 1178 | 8700 | 24 | 86.5% | 84.9% | 88.1% | 11.9% | 15.1% | 93.9% |
| **shadowing-local** | **10037** | 1819 | 2052 | 5597 | 569 | 74.7% | 76.2% | 73.2% | 26.8% | 23.8% | 81.9% |
| **variable-scope** | **10037** | 198 | 2308 | 7487 | 44 | 79.1% | 81.8% | 76.4% | 23.6% | 18.2% | 87.5% |
| **void-cst** | **10037** | 39 | 2303 | 7687 | 8 | 80.0% | 83.0% | 76.9% | 23.1% | 17.0% | 85.4% |

Table Appx.12: Task DLVA (SD + CC): use both the SD and CC; DLVA as a whole on $EthereumSC_{large}$

| Vulnerability | Test size | TP | FP | TN | FN | Accuracy | TPR | TNR | FPR | FNR |
|---|---|---|---|---|---|---|---|---|---|---|
| **shadowing-state** | **22634** | 578 | 1666 | 20238 | 152 | 85.9% | 81.5% | 90.3% | 9.7% | 18.5% |
| **suicidal** | **22634** | 76 | 1063 | 21487 | 8 | 91.9% | 89.6% | 94.0% | 6.0% | 10.4% |
| **uninitialized-state** | **22634** | 463 | 2592 | 19405 | 174 | 78.3% | 71.5% | 85.2% | 14.8% | 28.5% |
| **arbitrary-send** | **22634** | 1203 | 2461 | 18827 | 143 | 87.8% | 90.6% | 85.0% | 15.0% | 9.4% |
| **controlled-array-length** | **22634** | 905 | 1400 | 20199 | 130 | 90.0% | 88.4% | 91.7% | 8.3% | 11.6% |
| **controlled-delegatecall** | **22634** | 276 | 2087 | 20252 | 19 | 88.6% | 88.9% | 88.4% | 11.6% | 11.1% |
| **reentrancy-eth** | **22634** | 703 | 2377 | 19456 | 98 | 87.3% | 88.4% | 86.2% | 13.8% | 11.6% |
| **reentrancy-no-eth** | **22634** | 2556 | 2106 | 17575 | 397 | 86.6% | 87.4% | 85.9% | 14.1% | 12.6% |
| **unchecked-transfer** | **22634** | 2597 | 1400 | 18404 | 233 | 91.4% | 92.2% | 90.7% | 9.3% | 7.8% |
| **erc20-interface** | **22634** | 1613 | 1296 | 19592 | 133 | 91.2% | 90.1% | 92.3% | 7.7% | 9.9% |
| **incorrect-equality** | **22634** | 1473 | 2722 | 18142 | 297 | 84.3% | 85.8% | 82.8% | 17.2% | 14.2% |
| **locked-ether** | **22634** | 2351 | 2596 | 17489 | 198 | 85.3% | 86.0% | 84.6% | 15.4% | 14.0% |
| **mapping-deletion** | **22634** | 33 | 1336 | 21257 | 8 | 89.1% | 85.6% | 92.5% | 7.5% | 14.4% |
| **shadowing-abstract** | **22634** | 581 | 922 | 21089 | 42 | 93.0% | 91.3% | 94.8% | 5.2% | 8.7% |
| **tautology** | **22634** | 396 | 2928 | 19225 | 85 | 84.1% | 84.8% | 83.2% | 16.8% | 15.2% |
| **write-after-write** | **22634** | 72 | 3391 | 19146 | 25 | 79.5% | 77.9% | 81.1% | 18.9% | 22.1% |
| **constant-function-asm** | **22634** | 768 | 908 | 20910 | 48 | 94.6% | 94.4% | 94.8% | 5.2% | 5.6% |
| **constant-function-state** | **22634** | 39 | 2711 | 19879 | 5 | 87.2% | 89.4% | 84.9% | 15.1% | 10.6% |
| **divide-before-multiply** | **22634** | 2504 | 1826 | 17903 | 401 | 87.6% | 87.5% | 87.7% | 12.3% | 12.5% |
| **tx-origin** | **22634** | 55 | 2606 | 19966 | 7 | 88.3% | 91.2% | 85.5% | 14.5% | 8.8% |
| **unchecked-lowlevel** | **22634** | 285 | 925 | 21418 | 6 | 95.3% | 95.8% | 94.8% | 5.2% | 4.2% |
| **unchecked-send** | **22634** | 114 | 2387 | 20124 | 9 | 89.9% | 93.1% | 86.6% | 13.4% | 6.9% |
| **uninitialized-local** | **22634** | 1079 | 2709 | 18585 | 261 | 83.8% | 84.2% | 83.4% | 16.6% | 15.8% |
| **unused-return** | **22634** | 1940 | 2229 | 18116 | 349 | 86.4% | 87.5% | 85.3% | 14.7% | 12.5% |
| **incorrect-modifier** | **22634** | 213 | 1921 | 20463 | 37 | 87.3% | 85.5% | 89.2% | 10.8% | 14.5% |
| **shadowing-builtin** | **22634** | 286 | 1183 | 21136 | 29 | 91.8% | 90.2% | 93.4% | 6.6% | 9.8% |
| **shadowing-local** | **22634** | 4599 | 2201 | 15204 | 630 | 85.1% | 85.8% | 84.4% | 15.6% | 14.2% |
| **variable-scope** | **22634** | 253 | 2314 | 20018 | 49 | 86.5% | 86.2% | 86.9% | 13.1% | 13.8% |
| **void-cst** | **22634** | 56 | 2305 | 20261 | 12 | 84.7% | 82.1% | 87.1% | 12.9% | 17.9% |

Table Appx.13: DLVA on *EthereumSC$_{small}$* (only using DLVA's core classifier)

| Vulnerability | Test Size | TP | FP | TN | FN | Accuracy | TPR | TNR | FPR | FNR | AUC |
|---|---|---|---|---|---|---|---|---|---|---|---|
| **shadowing-state** | **1381** | 8 | 18 | 1355 | 0 | 98.7% | 100.0% | 98.7% | 1.3% | 0.0% | 94.7% |
| **suicidal** | **1381** | 9 | 143 | 1229 | 0 | 89.6% | 100.0% | 89.6% | 10.4% | 0.0% | 89.2% |
| **uninitialized-state** | **1381** | 10 | 32 | 1337 | 2 | 97.5% | 83.3% | 97.7% | 2.3% | 16.7% | 94.0% |
| **arbitrary-send** | **1381** | 54 | 16 | 1306 | 5 | 98.5% | 91.5% | 98.8% | 1.2% | 8.5% | 94.9% |
| **controlled-array-length** | **1381** | 8 | 20 | 1352 | 1 | 98.5% | 88.9% | 98.5% | 1.5% | 11.1% | 91.4% |
| **controlled-delegatecall** | **1381** | 5 | 25 | 1351 | 0 | 98.2% | 100.0% | 98.2% | 1.8% | 0.0% | 99.6% |
| **reentrancy-eth** | **1381** | 6 | 56 | 1319 | 0 | 95.9% | 100.0% | 95.9% | 4.1% | 0.0% | 99.6% |
| **reentrancy-no-eth** | **1381** | 18 | 2 | 1359 | 2 | 99.7% | 90.0% | 99.9% | 0.1% | 10.0% | 97.6% |
| **unchecked-transfer** | **1381** | 42 | 17 | 1318 | 4 | 98.5% | 91.3% | 98.7% | 1.3% | 8.7% | 96.4% |
| **erc20-interface** | **1381** | 22 | 64 | 1294 | 1 | 95.3% | 95.7% | 95.3% | 4.7% | 4.3% | 96.9% |
| **incorrect-equality** | **1381** | 19 | 26 | 1336 | 0 | 98.1% | 100.0% | 98.1% | 1.9% | 0.0% | 92.3% |
| **locked-ether** | **1381** | 71 | 37 | 1269 | 4 | 97.0% | 94.7% | 97.2% | 2.8% | 5.3% | 95.3% |
| **constant-function-asm** | **1381** | 5 | 11 | 1365 | 0 | 99.2% | 100.0% | 99.2% | 0.8% | 0.0% | 99.7% |
| **divide-before-multiply** | **1381** | 32 | 10 | 1336 | 3 | 99.1% | 91.4% | 99.3% | 0.7% | 8.6% | 95.5% |
| **unchecked-lowlevel** | **1381** | 11 | 4 | 1366 | 0 | 99.7% | 100.0% | 99.7% | 0.3% | 0.0% | 98.5% |
| **unchecked-send** | **1381** | 10 | 0 | 1371 | 0 | 100.0% | 100.0% | 100.0% | 0.0% | 0.0% | 100.0% |
| **uninitialized-local** | **1381** | 17 | 72 | 1290 | 2 | 94.6% | 89.5% | 94.7% | 5.3% | 10.5% | 95.6% |
| **unused-return** | **1381** | 473 | 3 | 894 | 11 | 99.0% | 97.7% | 99.7% | 0.3% | 2.3% | 99.2% |
| **incorrect-modifier** | **1381** | 36 | 6 | 1339 | 0 | 99.6% | 100.0% | 99.6% | 0.4% | 0.0% | 99.9% |
| **shadowing-builtin** | **1381** | 7 | 11 | 1363 | 0 | 99.2% | 100.0% | 99.2% | 0.8% | 0.0% | 99.7% |
| **shadowing-local** | **1381** | 27 | 90 | 1261 | 3 | 93.3% | 90.0% | 93.3% | 6.7% | 10.0% | 93.7% |